\documentclass[manuscript,screen,acmlarge,11pt]{acmart}

\AtBeginDocument{%
  \providecommand\BibTeX{{%
    \normalfont B\kern-0.5em{\scshape i\kern-0.25em b}\kern-0.8em\TeX}}}

\usepackage{booktabs} % For professional looking tables
\usepackage{multirow}
\usepackage{tabularx}
\usepackage{subcaption}
\usepackage{makecell}
\usepackage{framed}
\usepackage{enumitem}
\usepackage{xcolor}
\usepackage[most]{tcolorbox}
\usepackage{colortbl}
\usepackage{arydshln}
\usepackage[frozencache,cachedir=.]{minted}
\definecolor{dark-red}{RGB}{255,0,0}
\definecolor{dark-green}{RGB}{0,200,0}
\definecolor{grey}{gray}{0.5}

\newcommand{\chunk}[2]{%
	\fcolorbox{black}{yellow}{\bfseries\sffamily\scriptsize#1}%
   {$\blacktriangleright$#2$\blacktriangleleft$}%
}

\newcommand{\kisub}[1]{\chunk{Kisub}{{\textcolor{cyan}{\textsl{#1}}}}}

\newcommand{\xin}[1]{\textcolor{purple}{#1}}

\begin{document}

\title{Exploring Parameter-Efficient Fine-Tuning Techniques for Code Generation with Large Language Models}

\author{Martin Weyssow}
\authornote{Corresponding author.}
\email{martin.weyssow@umontreal.ca}
\affiliation{%
  \institution{DIRO, University of Montreal}
  \city{Montreal}
  \country{Canada}
}

\author{Xin Zhou}
\email{xinzhou.2020@phdcs.smu.edu.sg}
\affiliation{%
  \institution{Singapore Management University}
  \country{Singapore}
}

\author{Kisub Kim}
\email{falconlk00@gmail.com}
\affiliation{%
  \institution{Singapore Management University}
  \country{Singapore}
}

\author{David Lo}
\email{davidlo@smu.edu.sg}
\affiliation{%
  \institution{Singapore Management University}
  \country{Singapore}
}

\author{Houari Sahraoui}
\email{sahraouh@iro.umontreal.ca}
\affiliation{%
  \institution{DIRO, University of Montreal}
  \city{Montreal}
  \country{Canada}
}

% idk why this isn't showing
\authorsaddresses{%
Authors' addresses: Martin Weyssow, martin.weyssow@umontreal.ca, DIRO, University of Montreal, Canada; Xin Zhou, xinzhou.2020@phdcs.smu.edu.sg, Singapore Management University, Singapore; Kisub Kim, falconlk00@gmail.com, Singapore Management University, Singapore; David Lo, davidlo@smu.edu.sg, Singapore Management University, Singapore; Houari Sahraoui, sahraouh@iro.umontreal.ca, DIRO, University of Montreal, Canada.
}

\renewcommand{\shortauthors}{M. Weyssow et al.}

\begin{abstract}
Large language models (LLMs) demonstrate impressive capabilities to generate accurate code snippets given natural language intents in a zero-shot manner, \textit{i.e.}, without the need for specific fine-tuning.
While prior studies have highlighted the advantages of fine-tuning LLMs, this process incurs high computational costs, making it impractical in resource-scarce environments, particularly for models with billions of parameters.
To address these challenges, previous research explored in-context learning (ICL) and retrieval-augmented generation (RAG) as strategies to guide the LLM generative process with task-specific prompt examples. 
However, ICL and RAG introduce inconveniences, such as the need for designing contextually relevant prompts and the absence of learning task-specific parameters, thereby limiting downstream task performance.
In this context, we foresee parameter-efficient fine-tuning (PEFT) as a promising approach to efficiently specialize LLMs to task-specific data while maintaining reasonable resource consumption. 
In this paper, we deliver a comprehensive study of PEFT techniques for LLMs in the context of automated code generation.
Our comprehensive investigation of PEFT techniques for LLMs reveals their superiority and potential over ICL and RAG across a diverse set of LLMs and three representative Python code generation datasets: Conala, CodeAlpacaPy, and APPS.
Furthermore, our study highlights the potential for tuning larger LLMs and significant reductions in memory usage by combining PEFT with quantization.
Therefore, this study opens opportunities for broader applications of PEFT in software engineering scenarios. 
\end{abstract}





\maketitle

\section{Introduction}
\label{sec:introduction}

% P1. Trendy LLMs.. + representative examples.
Large Language Models (LLMs) based on the Transformer architecture~\cite{vaswani2017attention}, demonstrate significant potential in diverse domains, including natural language processing  (NLP)~\cite{kojima2022large,wei2022emergent,min2021recent}, computer vision~\cite{yang2023language,chappuis2022prompt,shao2023prompting}, and software engineering~\cite{chen2021evaluating,vaithilingam2022expectation,polycoder}. 
These models excel in generating high-quality content given natural language intents in zero-shot, \textit{i.e.}, without fine-tuning.
This capability has sparked considerable interest in the software engineering field for automating code-related tasks such as program repair~\cite{xia2022less, joshi2023repair, xia2023automated} and code generation~\cite{austin2021program,chen2021evaluating,nijkamp2023codegen}.

% P2. Full fine-tuning Effectiveness + Burden/Limitation.
While the zero-shot capabilities of LLMs are impressive, their full potential often emerges through fine-tuning~\cite{radford2019language,wei2021finetuned}.
Specifically, fine-tuning an LLM to task-specific data allows it to learn and encode knowledge of the potentially highly contextual data at hand and thus generate more meaningful content.
However, this process comes at a significant computational cost.
Full fine-tuning, where all the parameters of the LLMs are updated during training, demands remarkable computational resources, especially when the LLM contains billions of parameters~\cite{strubell2019energy}.
To mitigate this computational burden, prior studies in software engineering~\cite{xia2023automated, prenner2022can, zhou2023docprompting} have investigated prompt-engineering techniques such as In-Context Learning (ICL)~\cite{radford2019language, brown2020language} and Retrieval-Augmented Generation (RAG)~\cite{lewis2020retrieval}.
ICL consists of providing prompt examples of the task to the LLM, guiding it to generate contextually appropriate content without any fine-tuning involved. 
These examples can be manually created or randomly selected from a relevant training dataset.
This technique has already shown promising results for code-related tasks, including automated program repair~\cite{xia2023automated}, bug fixing~\cite{prenner2022can}, and code generation~\cite{zhou2023docprompting,wang2022execution,shrivastava2023repository}. 
Expanding upon ICL, RAG offers a more robust and powerful alternative that incorporate a knowledge retrieval system at inference.
Using RAG, a retrieval model fetches relevant information from an indexed corpus, such as code documentation or similar code snippets to the input problem. 
The retrieved information is then added to the input prompt to guide generation. 
Unlike ICL, which relies on pre-selected examples that may not always be tailored to the specific input, RAG dynamically adapts to each individual input problem, providing more relevant context.
This technique has demonstrated significant improvements in software engineering tasks such as code generation and summarization~\cite{parvez2021retrieval, liu2020retrieval, zhou2023docprompting}, code completion~\cite{lu2022reacc}, and program repair~\cite{wang2023rap}.

% P3. Limitations of ICL in terms of effectiveness.
Although ICL and RAG provide a viable alternative to full fine-tuning, it operates at inference time and does not involve learning task-specific parameters, which may prevent the LLM from capturing fine-grained information about the task and result in a loss of effectiveness.
In this context, Parameter-Efficient Fine-Tuning (PEFT) techniques have emerged as promising solutions to render the fine-tuning cost at the lowest while allowing the model to learn task-specific parameters. 
Prior works~\cite{choi2023codeprompt,wang2023one,wang2022no,saberi2024utilization} in code intelligence have demonstrated the capability of PEFT techniques, and often shown their superiority over full fine-tuning across a wide range of tasks.
However, these studies focus on small language models (SLMs) ($<$0.25B parameters) such as CodeBERT~\cite{feng2020codebert} and CodeT5~\cite{wang2021codet5} and overlooked the applicability of PEFT techniques to LLMs ($\geq$1B parameters), leaving an important research gap.
Given the growing ubiquity of LLMs, we believe addressing this gap is paramount in advancing the field of code intelligence and harnessing the full potential of LLMs.
Furthermore, we identify an additional research opportunity in exploring the usage of PEFT techniques under limited resource scenarios, aiming to demonstrate the democratization of LLMs tuning through PEFT.
Addressing these gaps will not only show how PEFT techniques can enhance the effectiveness of LLMs but also how they broaden the accessibility and utility of LLMs in scarce computation settings and alleviate the dependence of practitioners on large computational infrastructures.

% P4. In this paper, we delve into the behavior of LLMs under downstream tasks in the field of Software Engineering. 
In this paper, we present an empirical study on the usage of existing PEFT techniques with LLMs.
We focus our study on code generation, which has been a pivotal area of research due to its transformative impact on automating software development~\cite{chen2021evaluating,openai2023gpt,nijkamp2023codegen}.
Our objective is twofold. 
First, we aim to assess the code generation capabilities of LLMs using existing PEFT techniques such as LoRA~\cite{hu2021lora} and QLoRA~\cite{dettmers2023qlora} on datasets without test cases, including Conala~\cite{zhou2023docprompting} and CodeAlpacaPy~\cite{CodeAlpacaDataset}, as well as the APPS dataset~\cite{apps2021} with test cases. 
Second, we seek to compare the effectiveness of LLMs tuned with these PEFT techniques against SLMs, ICL, and RAG.
Additionally, we conduct our comparative study with limited availability of computational resources to investigate the broad practicality of using PEFT techniques for LLMs.
To achieve these objectives, we formulate four research questions that guide our study:

% P4-2. RQs
\begin{itemize}[leftmargin=*]
    \item[--] RQ1: How do LLMs and SLMs perform using ICL on the Conala and CodeAlpacaPy datasets?

    \item[--] RQ2: How do LLMs and SLMs perform using PEFT techniques on the Conala and CodeAlpacaPy datasets?

    \item[--] RQ3: How does LoRA compare with ICL and RAG on the Conala and CodeAlpacaPy datasets?

    \item[--] RQ4: Can we enhance the effectiveness of LLMs for code generation in the APPS dataset using LoRA and QLoRA?
\end{itemize}

Altogether, answering these four research questions fulfills both objectives of this empirical study. 
Our first three RQs focus on evaluating SLMs and LLMs for code generation on the Conala and CodeAlpaca datasets.
In RQ1, we illustrate the baseline effectiveness of SLMs and LLMs using ICL, which retrieves random examples from the training set to guide the model in generating code.
By addressing RQ2, we gain a comprehensive understanding of how effective SLMs and LLMs are when using different PEFT techniques.
In RQ3, we conduct a comparative study of the effectiveness of LoRA with ICL and RAG, a strong baseline that dynamically retrieves relevant examples by selecting those closest to the test instructions from the training set.
Finally, to showcase the potential broader impact of PEFT, we study in RQ4 whether tuning LLMs using LoRA and QLoRA can improve their effectiveness on APPS, a challenging benchmark with test cases.

% P5. Experiments and results
To address these RQs, we conduct experiments on three datasets, APPS~\cite{apps2021}, Conala~\cite{conala}, and CodeAlpacaPy specifically curated from CodeAlpaca~\cite{CodeAlpacaDataset} for Python code generation.
Conversely to evaluation datasets such as HumanEval~\cite{chen2021evaluating}, the APPS, Conala and CodeAlpaca datasets, widely used in prior code generation studies~\cite{zhou2023docprompting,wang2022execution,norouzi2021code, wang2022execution, wang2023codet5p, yuan2023evaluating}, include sufficient training examples that can be employed for fine-tuning.
For a comprehensive comparative analysis, we select four distinct model families:
CodeT5+~\cite{wang2023codet5p}, CodeGen~\cite{nijkamp2023codegen}, CodeGen2~\cite{nijkamp2023codegen2}, and CodeLlama~\cite{roziere2023code}, including eight large and three small variants.
Note that we omitted closed-sourced LLMs such as Codex due to the inaccessibility of their parameters, which makes the study of any fine-tuning technique infeasible.
Furthermore, our study incorporates six PEFT techniques: LoRA~\cite{hu2021lora}, IA3~\cite{liu2022few}, Prompt tuning~\cite{lester2021power}, and Prefix tuning~\cite{li2021prefix}.
In addition, we explore QLoRA~\cite{dettmers2023qlora} with 8-bit and 4-bit quantization, which combines LoRA and model quantization.
Unlike ICL and RAG, these techniques entail learning new parameters to tune the LLMs for the specific downstream task.
Our main findings are the following:
\begin{itemize}[leftmargin=*]
    % RQ1
    \item[--] ICL drastically improves the effectiveness of all models compared to a zero-shot prompt for code generation on Conala and CodeAlpacaPy.

    \item[--] Increasing the number of ICL examples does not always lead to improvement in effectiveness. Models achieve peak effectiveness with eight and four examples for Conala and CodeAlpacaPy, respectively.

    % RQ2
    \item[--] LLMs fine-tuned with LoRA, IA3, and Prompt tuning, \textit{i.e.,} a few millions of parameters, consistently outperform SLMs fully fine-tuned with hundreds of millions of parameters. 

    \item[--] Among PEFT techniques, LoRA achieves the highest effectiveness for the LLMs and SLMs.

    \item[--] QLoRA considerably reduces memory usage, achieving up to a 2-fold decrease compared to LoRA while improving or preserving the models' effectiveness. 
    Furthermore, QLoRA enables the fine-tuning of LLMs up to 34B parameters for less than 24GB of GPU memory.

    % RQ3
    \item[--] LoRA significantly enhances the performance of all models compared to ICL and RAG for code generation on Conala and CodeAlpacaPy.

    % RQ4
    \item[--] LoRA and QLoRA improve CodeLlama-7B-Instruct's effectiveness for code generation on the APPS dataset. 
\end{itemize}

\noindent Our study sheds light on the promising opportunities that PEFT techniques hold, warranting further exploration for their application in other code-related tasks and scenarios. 

% P6. Contributions
To summarize, our contributions are the following:
\begin{itemize}[leftmargin=*]
    \item[--] We conduct a comprehensive empirical study of six PEFT techniques, \textit{i.e.}, LoRA, IA3, Prompt tuning, Prefix tuning, QLoRA-8bit, and QLoRA-4bit, for Python code generation over a broad range of SLMs and LLMs.

    \item[--] A comprehensive comparison and analysis of PEFT techniques against ICL and RAG for LLMs on code generation. 

    \item[--] We demonstrate the practicality of leveraging PEFT techniques to effectively fine-tune LLMs of code and reduce the computational burden associated with full fine-tuning, showcasing their potential broader applications in software engineering.
\end{itemize}

\section{Background}
\label{sec:background}

\subsection{In-Context Learning (ICL) and Retrieval-Augmented Generation (RAG)}
As one of the specific types of LLM-related techniques, ICL has emerged as an effective technique~\cite{brown2020language,min2021metaicl,liang2022holistic,ouyang2022training, chowdhery2022palm}. 
ICL seeks to improve the abilities of LLMs by integrating context-specific information, in the form of an input prompt or instruction template, during the inference and thus without the need to perform gradient-based training. 
Therefore, by considering the context, the model becomes more capable of generating coherent and contextually relevant outputs. 
This contextual coherence of the LLM and not having to perform costly gradient-based training constitutes prime advantages of using ICL to specialize LLMs to a specific task or dataset. 
However, ICL also presents some inconveniences, including the need to design representative prompts~\cite{liu2022few, webson2021prompt, zhao2021calibrate}. 

RAG is a more sophisticated approach to inject examples into input prompts at inference.
Unlike ICL that select random examples, RAG relies on a retrieval model that dynamically retrieves examples from a dataset that are close to a query.
In practice, the query can be formulated using information from the test example at test time, such as the coding problem for the case of code generation.
Altogether, RAG allows injection more relevant information in the input prompt than ICL and has been succesfully applied to software engineering tasks, such as code generation~\cite{parvez2021retrieval, zhou2023docprompting}, code summarization~\cite{parvez2021retrieval, liu2020retrieval}, and code completion~\cite{lu2022reacc}.
Nonetheless, both ICL and RAG suffer from a few limitations.
One concerns the introduction of extra input tokens in the prompt, which may be infeasible when the contextual information is too large. 
Another limitation is the reliance on the quality and relevance of the retrieved examples. 
In RAG, the retrieval model must accurately find examples that are genuinely similar or useful for the test query. 
If the retrieval mechanism fails to identify appropriate examples, it can inject irrelevant or misleading information into the prompt, ultimately degrading performance.

\subsection{Parameter-Efficient Fine-Tuning (PEFT)}
%What PEFT is
PEFT refer to the utilization of techniques that optimize the fine-tuning process of LLMs by selectively updating a subset of parameters instead of updating the entire model's parameters~\cite{ding2022delta}.
Technically, PEFT techniques focus on learning a small number of parameters for the task at hand by designing additional layers~\cite{houlsby2019parameter}, adding prepending additional tokens~\cite{li2021prefix, lester2021power}, decomposing weight gradients into specific matrices~\cite{hu2021lora}.
One of the representative cutting-edge PEFT techniques is LOw-Rank Adaptation of LLMs (LoRA)~\cite{hu2021lora}.
The technique consists of freezing the model weights and injecting low-rank trainable matrices into the attention layers of the Transformer architecture~\cite{vaswani2017attention}, thereby drastically reducing the number of trainable parameters. 
We employ LoRA as one of our PEFT technique since it has been widely used in NLP~\cite{liu2022few,ding2022delta,treviso2023efficient} and showed promising performance.
We also employ IA3 which intends to improve upon LoRA and further reduces the amount of trainable parameters~\cite{liu2022few}.
In addition to LoRA and IA3, we also include Prompt tuning~\cite{lester2021power} and Prefix tuning~\cite{lester2021power} in our study.
Prompt tuning involves the process of prepending virtual tokens to the input tokens of the LLM, whereas Prefix tuning inserts virtual tokens in all the layers of the target model and thus requires learning more parameters.
These virtual tokens are differentiable, allowing them to be learned through backpropagation during fine-tuning, while the rest of the LLM remains frozen.
Furthermore, QLoRA~\cite{dettmers2023qlora} combines LoRA with model quantization, enabling the fine-tuning of LLMs with less GPU memory by reducing the precision of floating point data types within the model. 

\section{Applying LLMs with limited resources}
\label{sec:fine-tuning}

In the era of LLMs, the availability of substantial computational resources plays a crucial role in harnessing their high capabilities. 
Unfortunately, many researchers and practitioners often find themselves constrained by the limited availability of high-end computing infrastructures.

For instance, a software engineer with access to only a single consumer GPU (e.g., 24GB of VRAM) may find full fine-tuning impractical due to the significant memory demands. 
The rapid increase in model size and the number of trainable parameters exacerbates this issue. 
Despite its effectiveness, full fine-tuning comes at a steep computational cost~\cite{ding2023parameter,ouyang2022training,austin2021program}., underscoring a computation-effectiveness trade-off (see Table~\ref{tab:trade-off}).

To address these limitations, alternative approaches like ICL and RAG have gained attention. ICL and RAG offer a low-computation option by eliminating the need for parameter updates. 
However, these techniques comes with its own set of challenges, including the selection of representative examples and sensitivity to prompt design~\cite{liu2022few,webson2021prompt,zhao2021calibrate}.
In practice, this can result in lower effectiveness compared to fine-tuning, particularly for highly contextual tasks prevalent in software engineering.

\renewcommand{\arraystretch}{1}
\setlength{\arrayrulewidth}{.5pt}
\setlength{\tabcolsep}{3pt}
\begin{table}[!t]
\centering
\small
\caption{Computation-effectiveness trade-off for each model tuning technique.} 
\label{tab:trade-off}
\vspace{-1em}
    \begin{tabular*}{.5\linewidth}{l@{\extracolsep{\fill}}*{1}{rr}}
    \toprule
    Technique & Computation costs & Effectiveness \\
    \midrule
    Full fine-tuning & high [{\color{dark-red} X}] & high [{\color{dark-green} \checkmark}] \\
    ICL and RAG & low [{\color{dark-green} \checkmark}] & low [{\color{dark-red} X}] \\
    PEFT & low [{\color{dark-green} \checkmark}] & high [{\color{dark-green} \checkmark}] \\
    \bottomrule
    \end{tabular*}
\end{table}

\begin{figure}[!t]
    \centering
    \includegraphics[width=.7\textwidth]{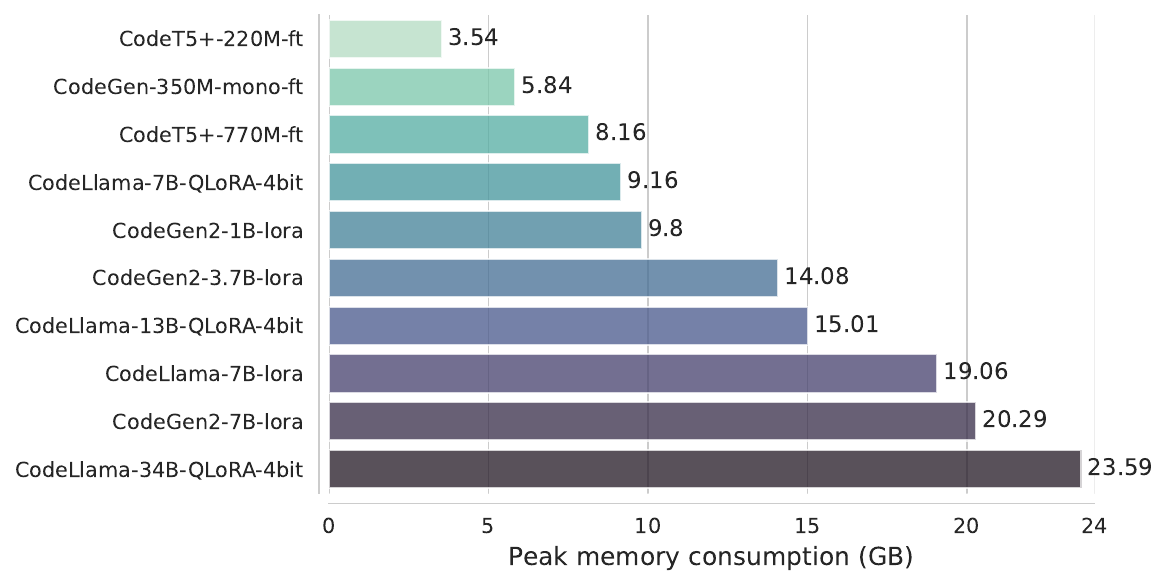}
    \vspace{-1em}
    \caption{Peak GPU memory consumption during models fine-tuning using full fine-tuning (ft), LoRA, and QLoRA.}
    \label{fig:gpu_memory_consumption}
\end{figure}

To overcome these limitations, we foresee the emergence of PEFT techniques as promising solutions, offering more computationally efficient and scalable approaches to fine-tuning LLMs. 
PEFT methods, such as LoRA and QLoRA, limit the number of parameters being updated, thus reducing memory consumption while maintaining effectiveness competitive with full fine-tuning. This makes PEFT particularly well-suited for practitioners with limited access to computational resources.
As illustrated in Table~\ref{tab:trade-off}, PEFT strike an optimal balance between computational cost and effectiveness. Furthermore, Fig.~\ref{fig:gpu_memory_consumption} shows that by employing PEFT techniques like LoRA, practitioners can fine-tune models such as CodeLlama-7B without exceeding 19GB of GPU memory. For even larger models, such as CodeLlama-34B, QLoRA with quantization enables fine-tuning within the constraints of a 24GB VRAM GPU.

In conclusion, PEFT empower software engineers to overcome resource limitations, allowing for effective LLM fine-tuning in highly contextual tasks without relying on expensive computational infrastructures. This makes PEFT not only a practical but also an essential tool for democratizing access to LLM capabilities.

\section{Methodology}

In this section, we present the experimental setup of our study. We conduct all the experiments under a resource-constrained scenario. Specifically, all the procedures, \textit{i.e.}, fine-tuning and inference, of the models are performed with access to a single 24GB GPU. The main objective of our study is to demonstrate whether the fine-tuning of LLMs through PEFT is feasible and desirable over previous approaches and smaller models in this context.

\subsection{Research Questions}
\label{sec:rqs}
In this study, we focus on the following research questions:
\begin{itemize}[leftmargin=*]
    \item[--] RQ1: \textbf{How do LLMs and SLMs perform using ICL on the Conala and CodeAlpacaPy datasets?}
    We study the baseline effectiveness of LLMs ($\geq$ 1B parameters) and SLMs ($<$ 1B parameters) for code generation using the zero-shot prompt and ICL, where $n$ randomly selected examples are added to the input prompt. We test each model with up to 16 ICL examples, due to our limited computation resources.
    
    We study the effectiveness of a large spectrum of SLMs and LLMs for code generation on two datasets covering codes of various lengths.
    We select a wide range of models of various sizes, pre-trained on diverse codebases and with different learning objectives to study how these factors impact their effectiveness. 

    \item[--] RQ2: \textbf{How do LLMs and SLMs perform using PEFT techniques on the Conala and CodeAlpacaPy datasets?}
    In this RQ, we investigate whether PEFT techniques consistently outperform ICL for SLMs and LLMs.
    We compare the best-performing configurations of ICL in RQ1 with PEFT techniques, including LoRA, IA3, Prompt tuning, Prefix tuning. 
    Furthermore, we also investigate the effect of quantization with QLoRA-8bit and QLoRA-4bit on our best-performing model and larger variants.
    
    For SLMs, we also include a comparison with full-parameter fine-tuning, as commonly used in previous SE studies~\cite{feng2020codebert,wang2021codet5,weyssow2022better,zhou2021assessing}. 
    We do not include full-parameter fine-tuning for LLMs, as it is not feasible within our computational budget.
     
    \item[--] RQ3: \textbf{How does LoRA compare with ICL and RAG on the Conala and CodeAlpacaPy datasets?}
    In this RQ, we compare the effectiveness of our best-performing LLM fine-tuned using LoRA with RAG. 
    Our RAG setup consists of retrieving up to $16$ examples from the training set that are closely related to the input prompt, which is similar to other approaches previously proposed for various SE tasks~\cite{liu2020retrieval, wang2023rap, lu2022reacc}.

    \item[--] RQ4: \textbf{Can we enhance the effectiveness of LLMs for code generation in the APPS dataset using LoRA and QLoRA?}
    Lastly, we explore whether LLM fine-tuned using LoRA and QLoRA show improvement in functional correctness in the APPS dataset.
    We fine-tune our best-performing LLM using LoRA and QLoRA on the training set of APPS, and report the average of test cases passed as well as the Pass@$k$ on APPS' test set for introductory, interview, and competition-level coding problems.
\end{itemize}

\subsection{Datasets and Task}
\label{sec:dataset_task}

\begin{figure}[!t]
    \centering
    \includegraphics[width=.6\linewidth]{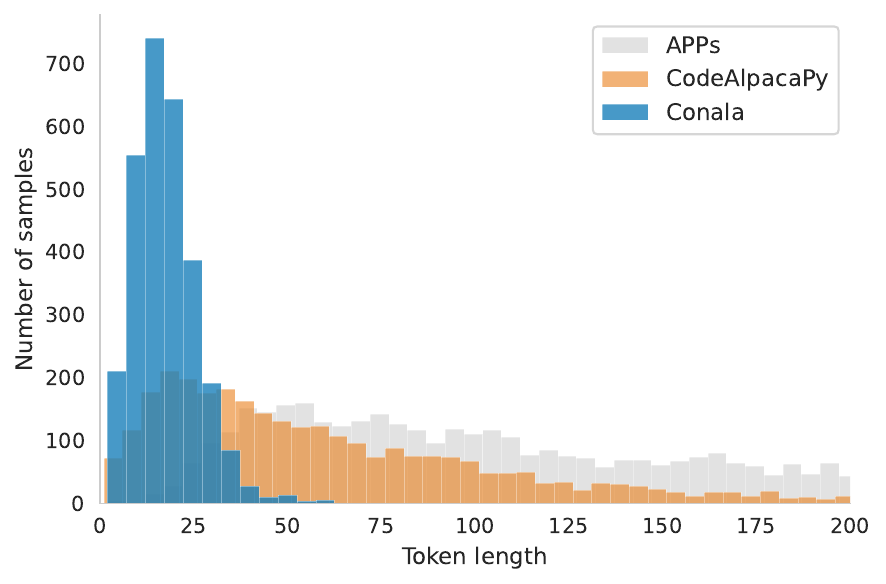}
    \vspace{-1em}
    \caption{Token length distribution of the Conala, CodeAlpacaPy, and APPS datasets.}
    \label{fig:dataset_dist}
\end{figure}

Throughout our study, we compare all the studied models on a Python code generation task. 
This task has gained significant attention in recent years~\cite{chen2021evaluating,openai2023gpt,nijkamp2023codegen,choi2023codeprompt,shrivastava2023repository} with the emergence of LLMs and their capability to generate Python code in zero-shot, \textit{i.e.,} without further fine-tuning. 
In particular, evaluation datasets such as HumanEval~\cite{chen2021evaluating} have extensively been used to benchmark code generation approaches~\cite{chen2021evaluating,austin2021program,polycoder}.
While HumanEval is widely utilized, it lacks a training corpus to evaluate fine-tuning or PEFT approaches. As our study's focus is on specializing LLMs using PEFT techniques, we have opted not to utilize HumanEval. 
Instead, we choose to use three other widely-used code generation datasets: the Conala~\cite{yin2018learning}, CodeAlpaca~\cite{CodeAlpacaDataset}, and APPS~\cite{apps2021} datasets.
All datasets provide an ample number of examples that can be employed for fine-tuning a model and have been used in prior code generation studies with LLMs~\cite{zhou2023docprompting,wang2022execution, wang2023codet5p, yuan2023evaluating}.

\subsubsection*{Conala dataset.}
We use a curated version of the Conala dataset \cite{zhou2023docprompting}.
The dataset was crawled from StackOverflow and contains manually annotated pairs of code and natural language intent. 
Each natural language intent contains hints about the manipulated variables in the ground truth code, \textit{e.g.}, see the first example in Table~\ref{tab:task_design}, providing more context to the model for generating relevant code.
In Figure~\ref{fig:dataset_dist}, we report the token length distributions of the three datasets. 
In Conala, most code solutions are short and one-liners, making it relatively easy for an LLM to generate exact match predictions.
In this curated version of the dataset, the authors ensured that each sample in the validation and test sets contained at least one Python function that does not appear in the training set. 
Additionally, they ensured that examples crawled from the same StackOverflow post appear in different sets.
Thus, we can guarantee that each natural intent in the test does not appear in the training set.
The dataset contains 2,135/201/543 samples as the training/validation/test sets, respectively.

\renewcommand{\arraystretch}{1}
\setlength{\arrayrulewidth}{.5pt}
\setlength{\tabcolsep}{3pt}
\begin{table}[!t]
\centering
\small
\caption{Overview of the code generation task, with three examples taken from the Conala, CodeAlpacaPy, and APPS datasets.} 
\label{tab:task_design}
\vspace{-1em}
    \begin{tabular*}{\linewidth}{ll}
    \toprule
    \multicolumn{2}{c}{\emph{Conala}} \\
    \midrule
    \arrayrulecolor{black!50}
    \underline{Prompt}: & {\makecell[l]{\#\#\# Instruction: \\ \textit{map two lists `keys' and `values' into a dictionary} \\ \#\#\# Response:}} \\
    \midrule
    \underline{Ground truth}: & \mintinline{python}{dict([(k, v) for k, v in zip(keys, values)])} \\
    \arrayrulecolor{black}
    \midrule
    \multicolumn{2}{c}{\emph{CodeAlpacaPy}} \\
    \midrule
    \arrayrulecolor{black!50}
    \underline{Prompt}: & {\makecell[l]{\#\#\# Instruction: \\ \textit{Write a function to calculate the standard deviation of data points in Python.} \\ \#\#\# Response:}} \\
    \midrule
    \underline{Ground truth}: & {\makecell[l]{
        \mintinline{python}{def stdev(data):}  \\
        \hspace{1em} \mintinline{python}{avg = sum(data) / len(data)} \\
        \hspace{1em} \mintinline{python}{total = 0} \\
        \hspace{1em} \mintinline{python}{for x in data:} \\
        \hspace{2em} \mintinline{python}{total += (x - avg) ** 2} \\
        \hspace{1em} \mintinline{python}{return (total / (len(data) - 1)) ** 0.5}
    }} \\
    \arrayrulecolor{black}
    \midrule
    \multicolumn{2}{c}{\emph{APPS}} \\
    \midrule
    \arrayrulecolor{black!50}
    \underline{Prompt}: & \parbox[c]{13.5cm}{\#\#\# Instruction: \\ \textit{You are given a string s = s1 s2 . . . sn of length n, which only contains digits 1, 2,..., 9.
    A substring s[l...r] of s is a string slsl+1sl+2 ...sr. A substring s[l...r] of s is called even if the number represented by it is even. 
    Find the number of even substrings of s. 
    Note, that even if some substrings are equal as strings, but have different l and r, 
    they are counted as different substrings. The first line contains an integer n 
    (1 $\leq$ n $\leq$ 65000) — the length of the string s. The second line contains a string s of length n. 
    The string s consists only of digits 1, 2,..., 9. Print the number of even substrings of s.} \\
    \#\#\# Response:} \\ 
    \midrule
    \underline{Ground truth}: & {\makecell[l]{
        \mintinline{python}{n = int(input())} \\
        \mintinline{python}{ans = 0} \\
        \mintinline{python}{for i in range(n):} \\
        \hspace{1em} \mintinline{python}{for j in range(i, n):} \\
        \hspace{2em} \mintinline{python}{if int(s[i:j+1]) \% 2 == 0:} \\
        \hspace{3em} \mintinline{python}{ans += 1} \\
        \mintinline{python}{print(ans)} \\    
    }} \\
    \arrayrulecolor{black}
    
    \bottomrule
    \end{tabular*}
\end{table}

\subsubsection*{CodeAlpacaPy dataset.} 
We construct a curated Python version of the CodeAlpaca~\cite{CodeAlpacaDataset} dataset by specifically selecting the Python data samples within the CodeAlpaca dataset.
We filter out code samples that cannot be statically parsed to ensure the dataset encompasses only syntactically valid Python codes.
As illustrated in the bottom example of Table~\ref{tab:task_design} and in Figure~\ref{fig:dataset_dist}, CodeAlpacaPy contains lengthier and more complex examples than Conala, allowing for a more comprehensive evaluation of PEFT for code generation.
The dataset contains 2,192/314/628 samples as the training/validation/test sets, respectively.

\subsubsection*{APPS dataset.} 
The APPS dataset consists of 10,000 code generation problems, each paired with Python solutions.
These problems are categorized into three difficulty levels: introductory, interview, and competition, with solutions varying from simple one-liners to complex algorithms. 
We can see in Fig.~\ref{fig:dataset_dist} and Table~\ref{tab:task_design} that APPS include more lengthy and complex examples than the two other datasets.
On average, each problem is accompanied by 21.2 test cases, designed to evaluate the functional correctness of the generated code. 
The original dataset is split into 5,000 samples for training and 5,000 for testing. 
In this study, we use 4,500 samples for training, 500 for validation, and 750 for testing, ensuring a balanced distribution of 250 test samples per difficulty level.

\subsubsection*{Task design} 
In Table~\ref{tab:task_design}, we illustrate an overview of the task design. 
The prompt is in the form of an instruction template, where ``\#\#\# Instruction:'' and ``\#\#\# Response:'' play the role of delimiting the instruction, \textit{i.e.}, natural language intent, and the answer, \textit{i.e.}, code generation. 
Note that this prompt design may not be optimal, but this kind of instruction template has shown to be effective in prior works~\cite{zhang2023investigating,liang2023gpt}.
The code generated by the model is compared with the ground truth to assess the quality of the generation. 
During fine-tuning, we minimize a standard autoregressive cross-entropy loss function:
$$
\mathcal{L} = - \sum_{i=1}^{T+1} M_i\cdot\log P(x_i\:|\:x_{<i})\:,
$$
where:
$$
M_i = \begin{cases} 
       1\:, & \text{if } x_i \neq -100 \\
       0\:, & \text{otherwise}.
      \end{cases}
$$
The model receives a concatenation of the prompt and the ground truth as input and predicts each token $x_i$ in an autoregressive manner given the previous tokens $x_{<i}$. 
Note that in the computation of the loss, we ignore the tokens from the instruction template to force the model to focus on generating code. 
We set the value of the instruction tokens to $-100$ and ignore them in the loss computation using the indicator function $M_i$.
At inference, the model receives the prompt as input and attempts to generate the ground truth code by generating up to 10 code candidates.

\subsection{ICL and RAG} 
We conduct experiments using ICL and RAG on the Conala and CodeAlpacaPy datasets. For both techniques, we select the maximum number of samples that can fit into our GPU memory.
For ICL, we use up to 16 examples for the Conala dataset and 8 examples for CodeAlpacaPy. These examples are randomly sampled from the corresponding training datasets and concatenated with the input prompt during inference.
For RAG, we leverage GTE-small, a general-purpose, lightweight embedding model that outperforms many larger models, including OpenAI's proprietary embeddings~\cite{li2023generaltextembeddingsmultistage}. We generate embeddings for all instructions (excluding the code) in the training sets. At inference time, we retrieve up to 16 examples for Conala and 4 examples for CodeAlpacaPy, selecting those with instructions most similar to the test input. As with ICL, the retrieved examples are concatenated with the input problem to guide code generation.

\subsection{Small and Large Language Models}
\label{sec:baselines_llms}
In order to carry out a comprehensive analysis, we selected our SLMs and LLMs according to several criteria.
First, we exclusively considered open-source models.
We omitted closed-sourced LLMs such as Codex due to the inaccessibility of their parameters, which makes the study of any fine-tuning technique infeasible.
All the studied models' checkpoints can be freely accessed, and have been pre-trained using open-source data.
Secondly, we selected LLMs, which have been released within the past two years. 
Finally, to investigate the impact of scaling, we selected models with a diverse range of parameters.
We consider models with less than 1B parameters as SLMs, and the others as LLMs.
Note that we selected models that fit a single 24GB GPU for fine-tuning and inference without causing memory overflow.
In total, we included 11 SLMs and LLMs from diverse families of models to conduct our experiments.

\begin{itemize}[leftmargin=*]
  \item[--] \textbf{SLMs}. We use CodeGen-350M-mono~\cite{nijkamp2023codegen}, CodeT5+-220M~\cite{wang2023codet5p}, and CodeT5+-770M~\cite{wang2023codet5p} as SLMs. 
  CodeGen-350M-mono is an autoregressive language model and a small version of CodeGen pre-trained on various programming languages and further fine-tuned on Python data.
  CodeT5+-220M and CodeT5+-770M are encoder-decoder language models that improve upon CodeT5 by leveraging a two-staged pre-training phase on natural language and code data, and new learning objectives. 

  \item[--] \textbf{CodeGen2}~\cite{nijkamp2023codegen2} is a family of prefix-based language models which combines the learning schemes of a bi-directional encoder and a uni-directional decoder. 
  CodeGen2 improves upon CodeGen~\cite{nijkamp2023codegen}, therefore we do not include the CodeGen family in our evaluation. 
  CodeGen2 models were pre-trained on a deduplicated version of TheStack~\cite{Kocetkov2022TheStack} spanning a wide range of languages. 
  We employ CodeGen2-1B, CodeGen2-3.7B and CodeGen2-7B.
  
  \item[--] \textbf{CodeLlama}~~\cite{roziere2023code} is a family of LLMs based on Llama 2~\cite{touvron2023llama}.
  Each model was initialized with Llama 2 and further pre-trained on code. 
  CodeLlama comes in three different variants: CodeLlama specialized for code, CodeLlama-Instruct specialized for instruction-tuning and CodeLlama-Python specialized for Python.
  We employ CodeLlama-7B, CodeLlama-7B-Instruct and CodeLlama-7B-Python to initiate our experiments. 
  In RQ4, we fine-tune CodeLlama-13B-Python and CodeLlama-34B-Python using QLoRA.

\end{itemize}

\subsection{Metrics}
\label{sec:metrics}

We measure the effectiveness of the models through widely used metrics in prior code generation work. 
For experiments on Conala and CodeAlpacaPy, we report the Exact Match (EM) and CodeBLEU~\cite{ren2020codebleu} metrics. 
Given a generated code and a ground truth, the EM returns 1 if both codes are identical, otherwise 0.
To evaluate the effectiveness of the models on a list of $k \in [1, 10]$ candidates, we report the EM@$k$, which computes the average correct predictions among a list of $k$ candidates. 
For our experiments on the APPS dataset, we report two metrics: the average number of test cases passed and Pass@$k$.
The average number of test cases passed evaluates how well the model performs by measuring the proportion of test cases that its generated code passes for each sample. In contrast, Pass@$k$ is a more stringent metric that measures the percentage of problems for which at least one of the top $k$ generated code samples passes all test cases, reflecting the model's ability to produce fully correct solutions within $k$ attempts.

\subsection{Implementation Details}
\label{sec:impl}

For all our experiments, we used a single NVIDIA RTX A5000 24GB GPU. 
We study a total of seven tuning techniques: Full fine-tuning, ICL, LoRA~\cite{hu2021lora}, IA3~\cite{liu2022few}, Prompt tuning~\cite{lester2021power}, Prefix tuning~\cite{li2021prefix}, and QLoRA~\cite{dettmers2023qlora}.
We implemented all the tuning techniques using HuggingFace~\cite{wolf2019huggingface} and PEFT~\cite{peft} libraries.

We used full fine-tuning only for the SLMs, as tuning all the parameters of the LLMs is computationally intractable within a maximum GPU memory of 24GB. 
We set the learning rate to $5e-5$.
For LoRA and IA3, we applied the low-rank matrix decomposition on the attention layers of the models and set $r = 16$ and $\alpha = 32$. For implementing QLoRA, we use 8-bit and 4-bit quantization~\cite{dettmers2022llmint8}. We set the learning rate to $3e-4$ for LoRA, IA3 and QLoRA.
For Prompt tuning and Prefix tuning, we prepended a set of $20$ trainable continuous virtual tokens to each input sample of the models and applied learning rates of $3e-3$ and $3e-2$.

We used Adafactor~\cite{shazeer2018adafactor} optimizer with 16-bit float precision for all models. 
We fine-tuned the models for a maximum of five epochs and evaluated them every $0.2 * len(train\_set)$ optimization steps.
We fine-tune all models with a batch size of 8. 
We selected the checkpoint with the lowest evaluation loss for inference and found that beam search with a beam size of $10$ yields the best effectiveness.
Given the various token length distribution and complexity of the datasets, we generate codes with up to 64, 128, and 1024 tokens for Conala, CodeAlpacaPy, and APPS, respectively.
We make our code publicly available: 
\url{https://github.com/martin-wey/peft-llm-code}.

\section{Experimental Results}
\label{sec:experimental-results}

%%%%%%%%%%%% RQ1 %%%%%%%%%%%%%%
\subsection{RQ1: Baseline Effectiveness of Models Using Zero-Shot and ICL}

\begin{figure}
    \centering
    \includegraphics[width=\linewidth]{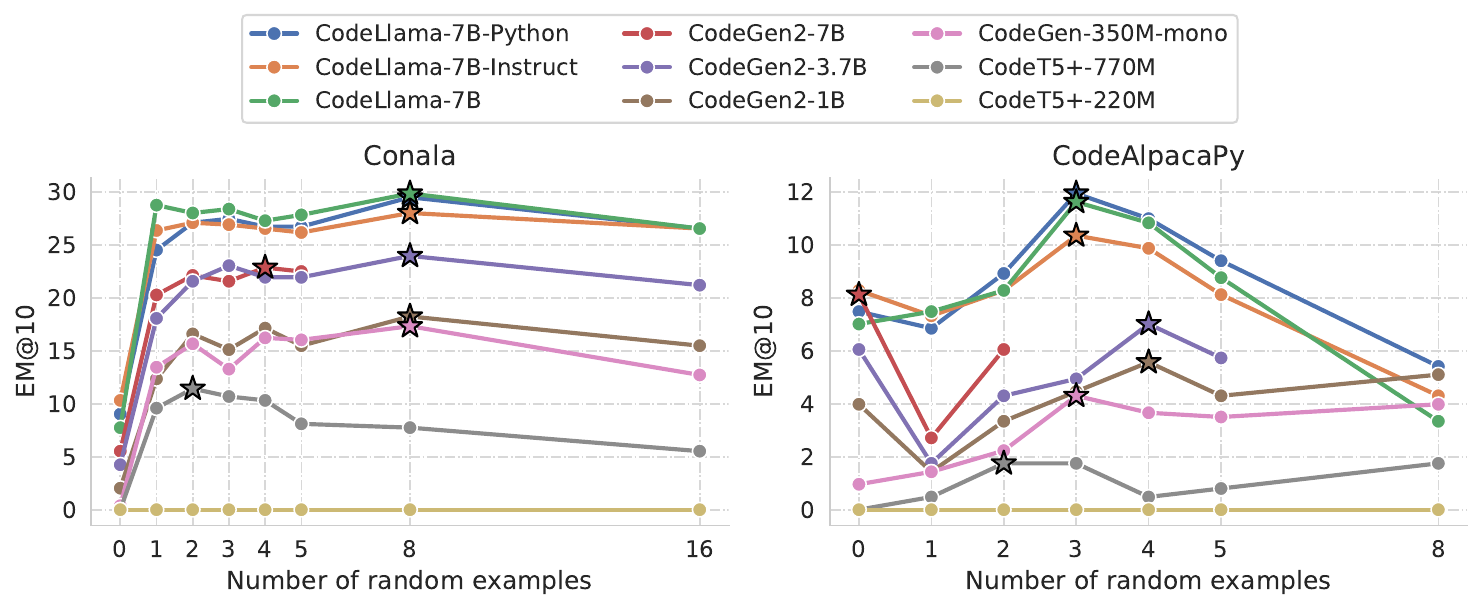}
    \vspace{-2em}
    \caption{[RQ1] -- Effectiveness of the models using ICL with various number of random examples on the Conala and CodeAlpacaPy datasets.}
    \label{fig:rq1_icl}
\end{figure}

We start by investigating the baseline effectiveness of all SLMs and LLMs for match-based code generation.
Specifically, we use zero-shot and ICL approaches with up to 16 retrieved random examples for the Conala dataset and eight for the CodeAlpacaPy dataset.
The reason behind utilizing fewer examples for CodeAlpacaPy is because considering 16 examples results in out-of-memory errors under our setup.
We evaluate the models' effectiveness using EM@10 and compare them across these two datasets in Fig.~\ref{fig:rq1_icl}.
Note that CodeGen2 architecture results in substantially more GPU memory usage than other models, which explains why we evaluate ICL with fewer examples than other models.

First, we observe a substantial gap in EM@10 between the two datasets. This difference can be explained by the fact that the CodeAlpacaPy dataset contains much more challenging samples compared to the Conala dataset, as shown in Table~\ref{tab:task_design}.

Second, there is a notable gap in effectiveness between SLMs and LLMs, regardless of the number of examples provided.
This observation highlights the advantages of large-scale pre-training and the use of larger models in this context.

For the Conala dataset, increasing the number of examples leads to higher EM@10 scores. However, when using more than eight examples, the effectiveness of the models begins to decline.
For the CodeAlpacaPy dataset, a similar trend is observed, but the optimal number of examples is smaller. Most models achieve their best EM@10 scores when using three or four examples.
This observation underscores the limitation of ICL, as adding more examples results in a degradation of the models' effectiveness.

Finally, CodeLlama models outperform all models across both datasets, achieving a peak EM@10 of 29.83 on Conala (CodeLlama-7B) and 11.94 on CodeAlpacaPy (CodeLlama-7B-Python). 
In contrast, smaller models, such as CodeGen2-3.7B achieves an EM@10 of 23.94 and 7.00 on Conala and CodeAlpacaPy, respectively.

\begin{tcolorbox}[tile,size=fbox,boxsep=2mm,boxrule=0pt,top=0pt,bottom=0pt,
borderline west={1mm}{0pt}{blue!50!white},colback=blue!10!white]
 \textbf{Answer to RQ1}: ICL drastically improves the effectiveness of all models compared to zero-shot. 
 Our best model, CodeLlama-7B, achieves EM@10 scores of 29.83 (7.73) and 11.62 (7.01) on Conala and CodeAlpacaPy with ICL (zero-shot), respectively.
\end{tcolorbox}

%%%%%%%%%%%% RQ2 %%%%%%%%%%%%%%
\subsection{RQ2: Effectiveness of Models using PEFT Techniques}

\renewcommand{\arraystretch}{1}
\setlength{\arrayrulewidth}{.5pt}
\setlength\extrarowheight{.5pt}
\setlength\dashlinedash{2pt}
\setlength\dashlinegap{2pt}
\begin{table}[!t]
\centering
\footnotesize
\caption{[RQ2] -- Comparison of the SLMs and LLMs using various tuning techniques (\colorbox{blue!25}{blue}: best-performing tuning method per model, \colorbox{orange!50}{orange}: best overall performing model).} 
\label{tab:rq2_table}
\vspace{-1em}
\resizebox{\columnwidth}{!}{%
    \begin{tabular}{llr*{6}{>{\centering\arraybackslash}p{1.5cm}}*{6}{c}}
    \toprule
    & & & \multicolumn{3}{c}{{\small \textbf{Conala}}} & \multicolumn{3}{c}{{\small \textbf{CodeAlpacaPy}}} \\ 
    \cmidrule(l{1em}r{1em}){4-6} \cmidrule(l{1em}r{1em}){7-9}
    \multicolumn{1}{c}{\textbf{{\small Model}}} & \multicolumn{1}{c}{{\small \textbf{Tuning}}} & \multicolumn{1}{c}{\small{\textbf{\# Params}}}  & \multicolumn{1}{c}{{\small EM@$1$}} & \multicolumn{1}{c}{{\small EM@$10$}} & \multicolumn{1}{c}{{\small CodeBLEU}} & \multicolumn{1}{c}{{\small EM@$1$}} & \multicolumn{1}{c}{{\small EM@$10$}} & \multicolumn{1}{c}{{\small CodeBLEU}} \\ 
    \midrule
    \multicolumn{9}{c}{\normalsize \textsc{SLMs}} \\
    \midrule
    \arrayrulecolor{black!50}

    \multirow{5}{*}{{\small CodeT5+-220M}} & Full FT & 220M & 3.87 & 8.84 & 16.70 & \cellcolor{blue!25} \textbf{3.98} & \cellcolor{blue!25} \textbf{7.64} & \cellcolor{blue!25} \textbf{25.49} \\

    & LoRA & 2.7M  & \cellcolor{blue!25} \textbf{6.08} & \cellcolor{blue!25} \textbf{12.71} & \cellcolor{blue!25} \textbf{18.96} & 3.34 &  5.26 & 24.32 \\

    & IA3 & 0.17M &  4.42 &  10.68 &  17.08 &  0.64 &  1.27 &  22.06 \\
    
    & Prompt tuning & 0.03M &  4.79 &  9.21 &  16.30 &  0.96 &  2.07 &  19.01 \\
    
    & Prefix tuning & 0.18M &  3.13 &  7.55 &  14.56 &  0.16 &  1.27 &  20.80 \\
    
    \cmidrule(l{1em}r{1em}){1-9}

    \multirow{5}{*}{{\small CodeT5+-770M}} & Full FT & 770M &  4.05 &  8.29 &  15.11 &  3.19 &  6.21 &  27.73 \\
    
    & LoRA & 7M & \cellcolor{blue!25} \textbf{8.66} &  17.13 & \cellcolor{blue!25} \textbf{20.64} & \cellcolor{blue!25} \textbf{3.66} & \cellcolor{blue!25} \textbf{6.85} & \cellcolor{blue!25} \textbf{26.10} \\
    
    & IA3 & 0.4M  &  8.10 & \cellcolor{blue!25} \textbf{17.50} &  18.68 &  2.87 &  5.26 &  25.84 \\
    
    & Prompt tuning & 0.04M &  7.37 &  15.47 &  16.75 &  1.91 &  3.82 &  20.57 \\
    
    & Prefix tuning & 0.5M &  4.97 &  11.97 &  16.77 &  0.16 &  1.27 &  22.91 \\
    
    \cmidrule(l{1em}r{1em}){1-9}

    \multirow{5}{*}{{\small CodeGen-350M-mono}} & Full FT & 350M &  7.92 &  18.42 &  14.68 &  2.23 &  5.73 &  21.78 \\
    
    & LoRA & 1.3M & \cellcolor{blue!25} \textbf{12.52} &  25.60 & 
     17.89 & \cellcolor{blue!25} \textbf{4.62} & \cellcolor{blue!25} \textbf{10.70} & \cellcolor{blue!25} \textbf{30.09} \\
    
    & IA3 & 0.16M &  11.42 & \cellcolor{blue!25} \textbf{25.78} & \cellcolor{blue!25} \textbf{18.83} &  4.46 & \cellcolor{blue!25} \textbf{10.70} &  28.56 \\
    
    & Prompt tuning & 0.02M &  7.92 &  20.26 &  16.29 &  0.0 &  0.0 &  25.91 \\
    
    & Prefix tuning & 0.4M &  5.34 &  12.52 &  17.53 &  0.0 &  0.0 &  26.89 \\

    \arrayrulecolor{black}
    \midrule
    \multicolumn{9}{c}{\normalsize \textsc{LLMs}} \\
    \midrule
    \arrayrulecolor{black!50}

    \multirow{4}{*}{{\small CodeGen2-1B}} & LoRA & 2M & 
     9.39 & \cellcolor{blue!25} \textbf{23.02} & \cellcolor{blue!25} \textbf{19.76} & \cellcolor{blue!25} \textbf{3.82} &  9.08 &  23.48 
    \\
    
    & IA3 & 0.2M &  10.13 &  22.84 &  18.64 & \cellcolor{blue!25} \textbf{3.82} & \cellcolor{blue!25} \textbf{9.87} & \cellcolor{blue!25} \textbf{24.42} \\
    
    & Prompt tuning & 0.04M & \cellcolor{blue!25} \textbf{11.97} &  22.65 &  18.38 &  0.80 &  2.07 &  18.17 \\
    
    & Prefix tuning & 0.6M &  5.89 &  15.84 &  18.46 &  0.0 &  0.32 &  13.68 \\

    \cmidrule(l{1em}r{1em}){1-9}
    \multirow{4}{*}{{\small CodeGen2-3.7B}} & LoRA & 4M & \cellcolor{blue!25} \textbf{11.60} &  25.97 &  19.00 & \cellcolor{blue!25} \textbf{5.41} &  10.70 &  23.75 \\
    
    & IA3 & 0.5M &  10.87 &  25.23 &  19.21 & \cellcolor{blue!25} \textbf{5.41} & \cellcolor{blue!25} \textbf{10.99} & \cellcolor{blue!25} \textbf{26.26} \\
    
    & Prompt tuning & 0.08M &  11.05 & \cellcolor{blue!25} \textbf{26.89} &  19.53 &  0.0 &  0.0 &  23.42 \\
    
    & Prefix tuning & 1.3M &  10.68 &  24.68 & \cellcolor{blue!25} \textbf{20.23} &  0.16 &  0.32 &  21.73 \\
    
    \cmidrule(l{1em}r{1em}){1-9}
    \multirow{4}{*}{{\small CodeGen2-7B}} & LoRA & 8.3M &  11.23 & \cellcolor{blue!25} \textbf{29.83} & \cellcolor{blue!25} \textbf{23.86} &  5.57 &  11.94 &  27.73 \\
    
    & IA3 & 1M &  11.42 &  29.65 &  21.98 & \cellcolor{blue!25} \textbf{5.73} & \cellcolor{blue!25} \textbf{12.42} & \cellcolor{blue!25} \textbf{28.26} \\
    
    & Prompt tuning & 0.08M & \cellcolor{blue!25} \textbf{11.97} &  27.26 &  22.37 &  0.0 &  0.0 &  25.40 \\
    
    & Prefix tuning & 2.6M &  9.95 &  23.94 &  22.29 &  0.0 &  0.32 &  25.72 \\

    \cmidrule(l{1em}r{1em}){1-9}
    \multirow{4}{*}{{\small CodeLlama-7B}} & LoRA & 12.5M & \cellcolor{orange!50} \textbf{20.07} & \cellcolor{orange!50} \textbf{39.31} & \cellcolor{blue!25} \textbf{25.33} &  7.33 & \cellcolor{blue!25} \textbf{16.24} & \cellcolor{blue!25} \textbf{32.05} \\
    
    & IA3 & 1M &  17.68 &  37.20 &  23.19 & \cellcolor{blue!25} \textbf{8.12} &  15.45 &  30.47 \\
    
    & Prompt tuning & 0.08M &  19.15 &  38.12 &  25.01 &  0.32 &  0.48 &  31.55 \\
    
    & Prefix tuning & 2.6M &  8.47 &  19.52 &  23.19 &  0.16 &  0.16 &  28.09 \\
    
    \cmidrule(l{1em}r{1em}){1-9}
    \multirow{4}{*}{{\small CodeLlama-7B-Instruct}} & LoRA & 12.5M &  17.68 & \cellcolor{blue!25} \textbf{36.28} &  24.27 &  7.01 & \cellcolor{orange!50} \textbf{17.04} & \cellcolor{blue!25} \textbf{31.42} \\
    
    & IA3 & 1M &  15.84 &  36.10 &  24.71 & \cellcolor{blue!25} \textbf{8.12} &  16.72 &  31.01 \\
    
    & Prompt tuning & 0.08M & \cellcolor{blue!25} \textbf{18.97} &  35.54 & \cellcolor{blue!25} \textbf{25.77} &  1.59 &  3.50 &  31.14 \\
    
    & Prefix tuning & 2.6M &  10.13 &  18.23 &  23.66 &  0.64 &  0.96 &  31.27 \\
    
    \cmidrule(l{1em}r{1em}){1-9}
    \multirow{4}{*}{{\small CodeLlama-7B-Python}} & LoRA & 12.5M & \cellcolor{blue!25} \textbf{17.50} &  36.28 &  24.27 &  7.96 &  15.92 &  32.84 \\
    
    & IA3 & 1M &  14.55 &  31.12 &  24.74 & \cellcolor{orange!50} \textbf{8.76} & \cellcolor{blue!25} \textbf{16.56} &  29.82 \\
    
    & Prompt tuning & 0.08M &  16.76 & \cellcolor{blue!25} \textbf{37.02} & \cellcolor{orange!50} \textbf{26.31} &  0.96 &  3.03 & \cellcolor{orange!50} \textbf{33.46} \\
    
    & Prefix tuning & 2.6M &  9.76 &  22.47 &  19.47 &  0.0 &  0.0 &  30.71 \\
    \arrayrulecolor{black}
    \bottomrule
    \end{tabular}
    }
\end{table}

We report the detailed results of the effectiveness of the SLMs and LLMs on match-based code generation for both Conala and CodeAlpacaPy datasets in Table~\ref{tab:rq2_table}. 

\subsubsection*{\textbf{SLMs vs. LLMs}} 
CodeGen-350M-mono with LoRA demonstrates the best effectiveness on average among small models, while CodeLlama-7B-Python with LoRA is the best LLM on average. 
Under the same 24GB GPU memory limitation, the best LLM surpasses the best small model by 39.8\%, 41.7\%, and 47.1\% (72.3\%, 48.8\%, and 9.1\%) in EM@1, EM@10, and CodeBLEU concerning the Conala (CodeAlpacaPy) dataset, respectively. % These notable results indicate a potential shift towards a new era of LLMs, differing from the popular paradigm in the SE community over the past 3-4 years: fine-tuning small pre-trained models such as CodeT5.

\subsubsection*{\textbf{SLMs}}
Among the SLMs, CodeGen-350M-mono shows the highest effectiveness across all metrics on both datasets. 
Our results align with prior studies~\cite{zhou2023docprompting, wang2022execution, nijkamp2023codegen} that identified CodeGen-350M-mono as a robust SLM for Python code generation tasks. Interestingly, although it requires tuning approximately 1\% of the total parameters of the model, LoRA appears as the best tuning technique, surpassing full fine-tuning by a considerable margin across nearly all configurations. 
For instance, the EM@10 score for CodeGen-350M-mono on the Conala dataset, with full fine-tuning, is 18.42, while it soars to 25.60 with LoRA.

\subsubsection*{\textbf{LLMs}} 
\begin{figure}[t!]
\centering
\begin{subfigure}[t]{.5\textwidth}
  \centering
  \includegraphics[width=\linewidth]{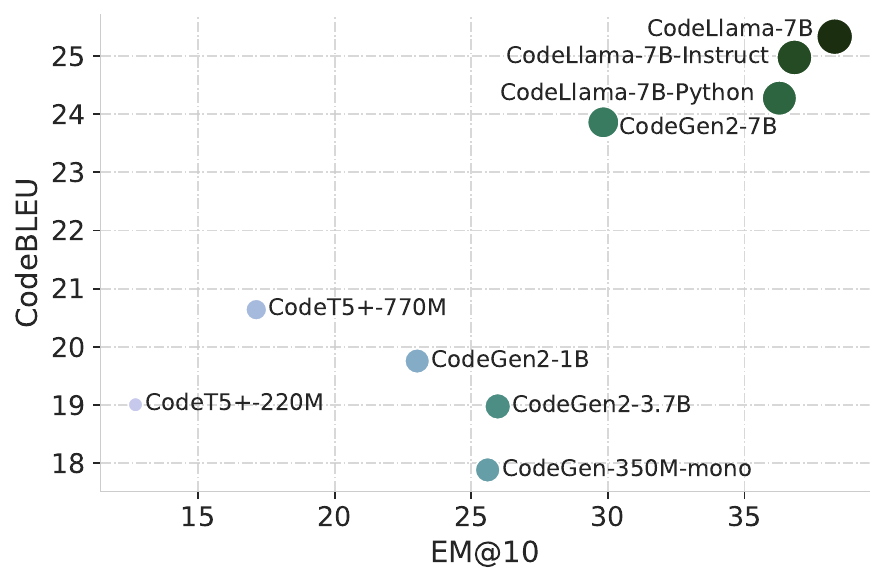}
  \caption{Conala}
  \label{fig:rq1_scatter_Conala}
\end{subfigure}%
\begin{subfigure}[t]{.5\textwidth}
  \centering
  \includegraphics[width=\linewidth]{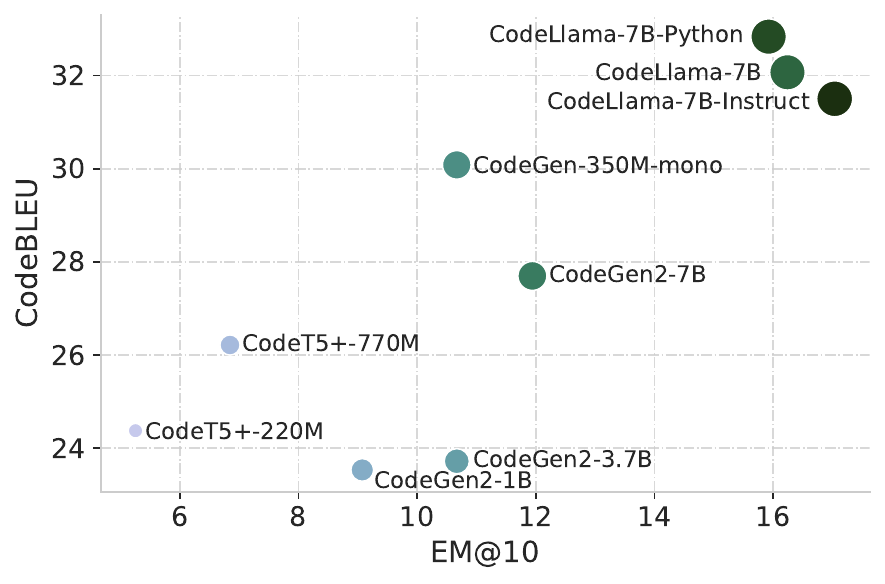}
  \caption{CodeAlpacaPy}
  \label{fig:rq1_scatter_codealpaca}
\end{subfigure}
\vspace{-1em}
\caption{[RQ2] -- Effectiveness of the models fine-tuned using LoRA for both datasets in terms of EM@10 and CodeBLEU.}
\label{fig:rq2_scatter_plots}
\end{figure}

In Figure~\ref{fig:rq2_scatter_plots}, we present a comparative analysis of the models' effectiveness when tuned using LoRA, focusing on CodeBLEU and EM@10 scores. 
Both plots clearly establish CodeLlama models as the best-performing LLMs in our study.
Remarkably, CodeGen2-7B, despite sharing a similar number of parameters, lags behind all CodeLlama-7B variants.
Unsurprisingly, harnessing larger models leads to better effectiveness.
Given the low computational costs of PEFT techniques, leveraging smaller models in a context akin to ours seems counterproductive.
Subsequently, in this paper, we demonstrate that even larger models can be fine-tuned through the combination of PEFT with quantization.

\subsubsection*{\textbf{Best PEFT technique}}
Overall, LoRA emerges as the most effective PEFT technique among the studied ones. 
Although being presented as an incremental improvement over LoRA~\cite{liu2022few}, IA3 often shows lower scores compared to LoRA.
Prompt tuning appears as another viable tuning option, while further reducing the number of trainable parameters. 
However, Prefix tuning fails to effectively adapt the larger models to both datasets.

Our analysis reveals notably higher EM scores for the Conala dataset, which can be attributed to differences in task complexity between the two datasets (see Section~\ref{sec:dataset_task}).
It is important to note that CodeBLEU scores on Conala are comparatively lower due to the metric's reliance on dataflow graph computations, which may not always be available for small code examples.

\subsubsection*{\textbf{Effect of quantization with QLoRA}}
\begin{figure*}[!t]
    \centering
    \includegraphics[width=.85\linewidth]{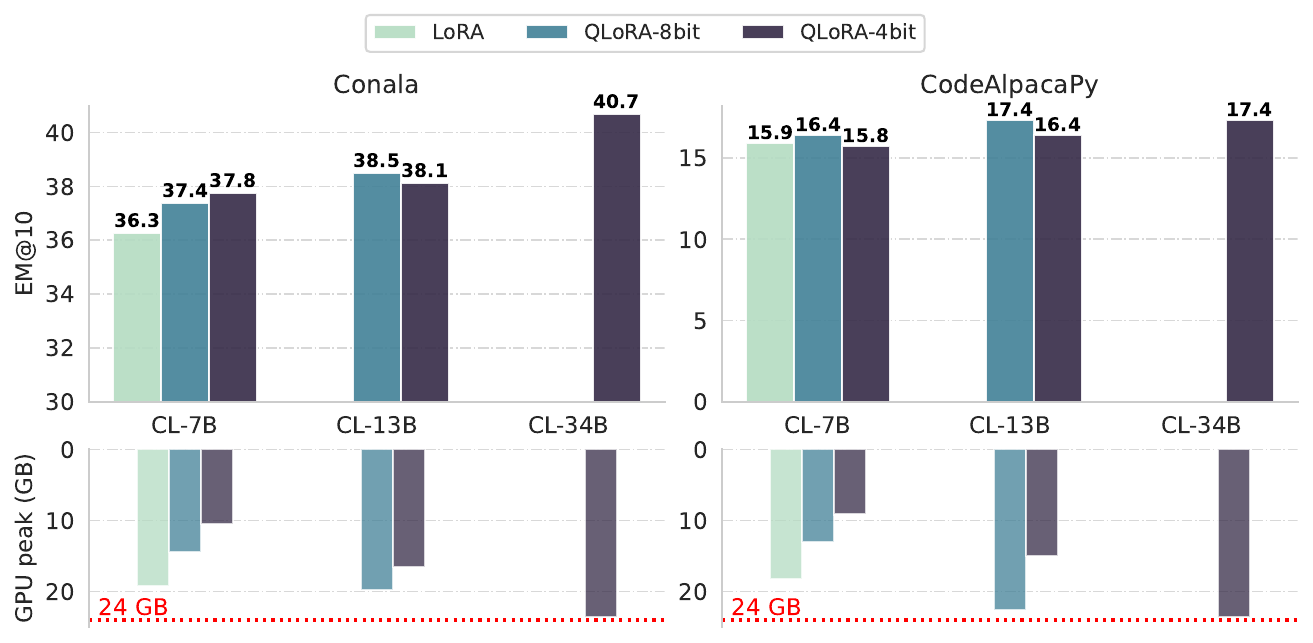}
    \vspace{-1em}
    \caption{[RQ2] -- Effectiveness and GPU usage of 7B, 13B, and 34B CodeLlama-Python (CL) LLMs fine-tuned using LoRA and QLoRA with 8-bit and 4-bit quantization.
    }
    \label{fig:rq2_qlora}
\end{figure*}

We explore the potential benefits of employing QLoRA~\cite{dettmers2023qlora}, a computationally efficient technique that combines LoRA with 8-bit or 4-bit quantization for fine-tuning LLMs. 
In Figure~\ref{fig:rq2_qlora}, we display EM@10 scores for three CodeLlama model variants: CodeLlama-7B-Python, CodeLlama-13B-Python, and CodeLlama-34B-Python, alongside peak GPU memory consumption consistently below 24GB for each tuning configuration.
The results underscore a significant improvement in the effectiveness of larger quantized models on Conala, with a more moderate impact on CodeAlpacaPy.
For instance, CodeLlama-34B-Python, fine-tuned with QLoRA-4bit, achieves a substantial 12.2\% increase in Conala's EM@10 score (40.70) compared to CodeLlama-7B-Python with LoRA (36.28). 
Surprisingly, QLoRA also brings notable improvements over LoRA for CodeLlama-7B-Python on Conala, while achieving comparable results on CodeAlpacaPy.
The application of quantization enables the utilization of larger models that can be accommodated within a single 24GB GPU.
Specifically, for CodeLlama-7B-Python, QLoRA-4bit achieves a remarkable 2x reduction in peak memory usage while significantly improving the EM@10 score.

\begin{tcolorbox}[tile,size=fbox,boxsep=2mm,boxrule=0pt,top=0pt,bottom=0pt,
borderline west={1mm}{0pt}{blue!50!white},colback=blue!10!white]
 \textbf{Answer to RQ2}: LLMs with PEFT consistently and significantly outperform SLMs under the same GPU limit.
 Specifically, the best-performing LLM with PEFT surpasses the best small model by 39.8--72.3\% in terms of EM@$k$.
 Among different PEFT techniques, LoRA is the most effective. In addition, applying quantization with LoRA results in a drastic decrease in GPU usage while maintaining effectiveness on both datasets and accommodating the fine-tuning of larger models up to 34B parameters.
\end{tcolorbox}

%%%%%%%%%%%% RQ3 %%%%%%%%%%%%%%
\subsection{RQ3: Comparative Analysis of LoRA, ICL, and RAG}

In this RQ, we aim to investigate whether PEFT techniques consistently outperform the widely used ICL and RAG when applying LLMs in match-based code generation.

\begin{figure}
    \centering
    \includegraphics[width=0.8\linewidth]{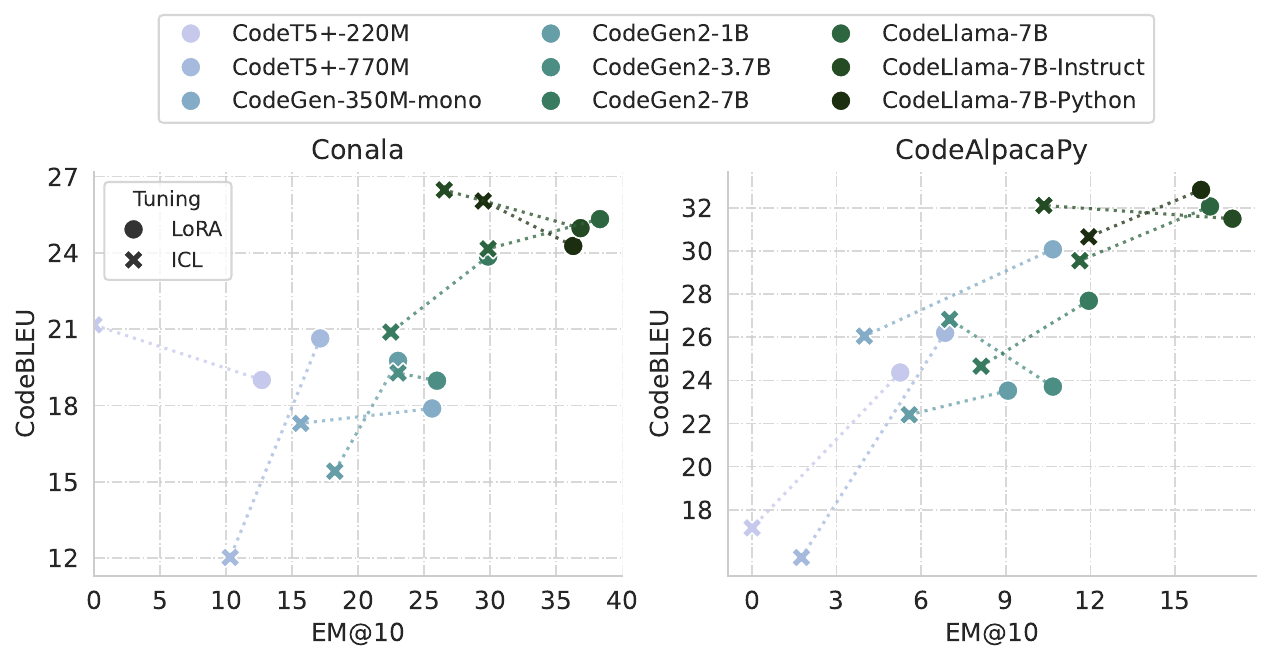}
    \vspace{-1em}
    \caption{[RQ3] -- Comparison of the effectiveness of the models fine-tuned using LoRA and ICL on the Conala and CodeAlpacaPy datasets.}
    \label{fig:rq3_lora_vs_icl}
\end{figure}

\begin{figure}
    \centering
    \includegraphics[width=0.75\linewidth]{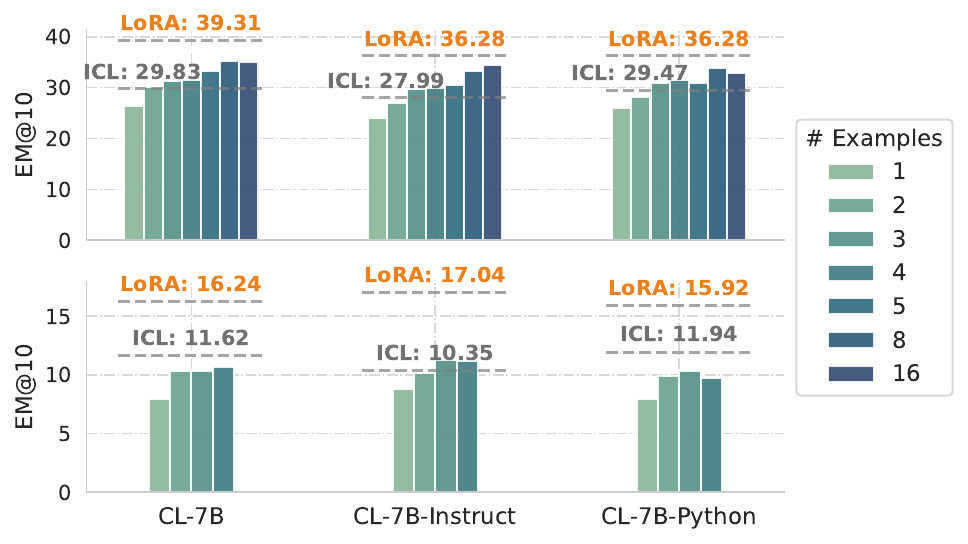}
    \vspace{-1em}
    \caption{[RQ3] -- Comparison of the effectiveness of RAG with various number of retrieved examples against ICL and LoRA on the Conala (top) and CodeAlpacaPy (bottom) datasets. The ICL scores depict the highest scores achieved for each model in RQ1.}
    \label{fig:rq3_rag}
\end{figure}

In Figure~\ref{fig:rq3_lora_vs_icl}, we compare the effectiveness of the SLMs and LLMs using ICL and LoRA in terms of CodeBLEU and EM@10.
In this figure, we report the highest metrics achieved over the different ICL configurations for each model.
In Figure~\ref{fig:rq3_rag}, we explore the effectiveness of CodeLlama models using RAG, with up to 16 and 4 retrieved examples for Conala and CodeAlpacaPy, respectively.
Similar to RQ1, we use fewer examples for CodeAlpacaPy to avoid out-of-memory errors.
We compare the effectiveness of RAG with LoRA and the best EM@10 score achieved using ICL.

\subsubsection*{\textbf{LoRA vs. ICL}}
As shown in Fig.~\ref{fig:rq3_lora_vs_icl}, all models fine-tuned with LoRA demonstrate significantly higher EM@10 scores compared to ICL across both datasets.
For example, CodeLlama-7B-Python with LoRA tuning achieves a 23.1\% improvement in EM@10 on Conala (36.28 for LoRA vs. 29.47 for ICL).
This pattern holds for CodeAlpacaPy, with even greater relative gains in EM@10.
However, we observe some variation in CodeBLEU scores for most models on CodeAlpacaPy. For instance, CodeLlama-7B sees a CodeBLEU increase of 2.36 with LoRA. On CoNala, though, the impact of LoRA on CodeBLEU is less pronounced than that of ICL.
These differences can be explained by the nature of the metrics: EM@10 is more conservative, requiring the generated solution to exactly match the ground truth, while CodeBLEU gives higher scores for solutions that are close but not exact.
This distinction highlights how LoRA better adapts models to downstream datasets, particularly when precision is crucial.

\subsubsection*{\textbf{RAG vs. ICL vs. LoRA}} 
In comparing RAG, ICL, and LoRA on the CoNala dataset, RAG demonstrates higher effectiveness than ICL but falls short of LoRA's effectiveness across all three CodeLlama model variants.
Notably, CodeLlama-7B achieves a maximum of 29.83 and 35.17 EM@10 with ICL and RAG, respectively, whereas the model tuned with LoRA reaches an EM@10 of 39.31.

For both Conala and CodeAlpacaPy datasets, the gains in EM@10 get thinner as we increase the number of examples using RAG. 
EM@10 saturates at around 8–16 examples for Conala and 3–4 examples for CodeAlpacaPy.
Furthermore, we note that for the more challenging CodeAlpacaPy datasets, RAG yields lower EM@10 compared to randomly selected examples using ICL, highlighting RAG's limitations when problem complexity increases. 
LoRA, however, consistently outperforms both RAG and ICL on CodeAlpacaPy, highlighting its superior ability to adapt to more challenging datasets.

\begin{tcolorbox}[tile,size=fbox,boxsep=2mm,boxrule=0pt,top=0pt,bottom=0pt,
borderline west={1mm}{0pt}{blue!50!white},colback=blue!10!white]
 \textbf{Answer to RQ3}: LoRA is superior to ICL and RAG on Conala and CodeAlpacaPy datasets across the three CodeLlama-7B variants.
\end{tcolorbox}

%%%%%%%%%%%% RQ4 %%%%%%%%%%%%%%
\subsection{RQ4: Exploration of LoRA and QLoRA for Code Generation on APPs}

In this final RQ, we explore the broader applicability of LoRA and QLoRA, to enhance CodeLlama-7B-Instruct's effectiveness for execution-based code generation.
The reason for choosing the instruct variant of CodeLlama-7B is because the model generally shows higher effectiveness than the other model variants on APPs in the seminal paper of CodeLlama~\cite{roziere2023code}.
We do not compare LoRA and QLoRA with ICL and RAG for this dataset because they require increasing the prompt length beyond 2,048 tokens, which leads to out-of-memory errors.
Our results, summarized in Table~\ref{tab:rq4}, focus on the average number of test cases passed (Avg) and Pass@$k$ for introductory, interview, and competition-level tasks.

For both introductory and interview-level code generation tasks, LoRA and QLoRA-8/4bit lead to significant improvements in the average number of passed test cases. 
Specifically, QLoRA-4bit results in a notable 52\% increase in the average number of tests passed compared to the base model.
In terms of Pass@$k$ metrics, both LoRA and QLoRA-4bit demonstrate gains at the introductory level, with Pass@5 improving by +3.60\% over the base model.
However, these improvements are less substantial for interview and competition-level code generation, reflecting the greater complexity and challenge posed by these more advanced tasks.

\renewcommand{\arraystretch}{1}
\setlength{\arrayrulewidth}{.5pt}
\setlength\extrarowheight{.5pt}
\begin{table}[!t]
\centering
\footnotesize
\caption{[RQ4] -- Effectiveness of CodeLlama-7B-Instruct on the APPs dataset in zero-shot and using LoRA, QLoRA-8bit, and QLoRA-4bit in terms of average passed tests (Avg) and Pass@k (P@k).} 
\label{tab:rq4}
\vspace{-1em}
\resizebox{\columnwidth}{!}{%
    \begin{tabular}{l*{12}{>{\centering\arraybackslash}p{.8cm}}*{12}{c}}
    \toprule
    & \multicolumn{4}{c}{{\small \textbf{Introductory}}} & \multicolumn{4}{c}{{\small \textbf{Interview}}} & \multicolumn{4}{c}{{\small \textbf{Competition}}} \\ 
    \cmidrule(l{1em}r{1em}){2-5} \cmidrule(l{1em}r{1em}){6-9} \cmidrule(l{1em}r{1em}){10-13} 
    \textbf{\small Model} & \multicolumn{1}{c}{{\small Avg}} & \multicolumn{1}{c}{{\small P@1}} & \multicolumn{1}{c}{{\small P@2}} & \multicolumn{1}{c}{{\small P@5}} & \multicolumn{1}{c}{{\small Avg}} & \multicolumn{1}{c}{{\small P@1}} & \multicolumn{1}{c}{{\small P@2}} & \multicolumn{1}{c}{{\small P@5}} & \multicolumn{1}{c}{{\small Avg}} & \multicolumn{1}{c}{{\small P@1}} & \multicolumn{1}{c}{{\small P@2}} & \multicolumn{1}{c}{{\small P@5}} \\
    \midrule
    \arrayrulecolor{black!50}
    CodeLlama-7B-Instruct & 13.66 & 4.16 & 6.24 & 8.80 & 13.44 & 0.80 & 1.32 & 2.40 & 6.27 & \cellcolor{orange!50} \textbf{0.56} & \cellcolor{orange!50} \textbf{1.00} & \cellcolor{orange!50} \textbf{2.00} \\ 
    \midrule
    \hspace{1em} +LoRA & 19.57 & 5.60 & 8.04 & 11.20 & 16.96 & \cellcolor{orange!50} \textbf{1.04} & \cellcolor{orange!50} \textbf{1.80} & \cellcolor{orange!50} \textbf{3.20} & 6.93 & 0.32 & 0.60 & 1.20 \\
    \midrule
    \hspace{1em} +QLoRA-8bit & 17.63 & 3.68 & 5.40 & 7.60 & 15.53 & \cellcolor{orange!50} \textbf{1.04} & 1.64 & 2.40 & \cellcolor{orange!50} \textbf{7.59} & 0.24 & 0.48 & 1.20 \\
    \midrule
    \hspace{1em} +QLoRA-4bit & \cellcolor{orange!50} \textbf{20.84} & \cellcolor{orange!50} \textbf{5.76} & \cellcolor{orange!50} \textbf{8.40} & \cellcolor{orange!50} \textbf{12.40} & \cellcolor{orange!50} \textbf{20.34} & \cellcolor{orange!50} \textbf{1.04} & 1.76 & 2.80 & 6.66 & 0.48 & 0.88 & 1.60 \\
    \arrayrulecolor{black}
    \bottomrule
    
    \end{tabular}
    }
\end{table}

\begin{tcolorbox}[tile,size=fbox,boxsep=2mm,boxrule=0pt,top=0pt,bottom=0pt,
borderline west={1mm}{0pt}{blue!50!white},colback=blue!10!white]
 \textbf{Answer to RQ4}: 
LoRA and QLoRA enhance CodeLlama-7B-Instruct's effectiveness on APPs, particularly at the introductory, with QLoRA-4bit boosting the average number of passed test cases by 52\% and Pass@5 by 40\%.
However, improvements are less notable for interview and competition-level tasks.
\end{tcolorbox}

\section{Discussion}
\label{sec:discussion}
Our study explores PEFTs applied to code LLMs, elucidating the positive impact of these applications in efficiently tuning LLMs to task-specific datasets for code generation. 
In particular, our study illustrates the practicality of fine-tuning LLMs using PEFT, thereby alleviating the dependence of practitioners on large and expensive infrastructures.
Our findings also pinpoint several promising areas for future exploration, including the investigation of efficient techniques across diverse fine-tuning settings, during inference, and for other SE tasks.

\subsubsection*{\textbf{Efficient techniques for LLMs of code}}
Our work emphasizes efficient fine-tuning techniques, democratizing the tuning of LLMs to a broad audience. 
Nonetheless, our study did not include the exploration of efficient techniques for low-cost inference.
While PEFT techniques require additional fine-tuning time compared to ICL and RAG, it is noteworthy that these techniques do not impose any supplementary time cost during inference.
Nonetheless, we acknowledge the necessity of future investigations into techniques to reduce the time cost associated with LLMs during inference.

PEFT and ICL/RAG are non-exclusive techniques that can be used jointly. 
However, we decided not to include experiments on the application of ICL/RAG to LLMs fine-tuned using PEFT.
In practice, increasing the number of ICL/RAG examples at inference entails increased computational overhead as the token length of the prompt expands.
Consequently, we contend that employing ICL/RAG on a fine-tuned LLM might counterproductively escalate computational demands, outweighing potential benefits.

From a different angle, prior studies~\cite{yadav2023exploring,weyssow2023usage,gao2023keeping} highlighted the need to consider pre-trained language models and LLMs of code in continual learning settings. 
In this paradigm, the model must dynamically adapt to new data over time while preserving performance on previously seen data. 
In the specific setting of continuously evolving LLMs, PEFT techniques potentially offer valuable benefits. 
Nonetheless, it is yet to be determined whether PEFT techniques can efficiently adapt LLMs under a continual learning setting for code-related tasks, without compromising the retention of past knowledge.

\subsubsection*{\textbf{Effectiveness of QLoRA}}
Across all study datasets, we observed that QLoRA-4bit demonstrated competitive or comparable effectiveness to other PEFT methods. 
Notably, QLoRA-4bit outperformed LoRA and QLoRA-8bit on the Conala and APPs datasets. 
We hypothesize that this improvement stems from the regularization effect of reducing weight precision to 4 bits, which helps stabilize fine-tuning and mitigates overfitting. These findings highlight the potential for more efficient PEFT techniques, though further exploration is needed to fully understand their broader applicability.

\subsubsection*{\textbf{New findings for PEFT in software engineering}}
Our findings in RQ1 reveal that PEFT methods outperform full fine-tuning for SLMs in code generation tasks. 
This stands in contrast to prior large-scale studies in NLP, such as Ding et al.~\cite{ding2022delta}, which demonstrated the superior effectiveness of full fine-tuning over techniques like LoRA, Prompt Tuning, and Prefix Tuning across a wide range of NLP tasks.

In the context of software engineering, while previous studies \cite{liu2024delving, liu2023empirical} have shown that PEFT methods, like LoRA, can perform comparably to full fine-tuning for SLMs, our results go further. 
We show that all PEFT techniques studied in this paper significantly outperform full fine-tuning for SLMs like CodeGen-350M-mono and CodeT5+-770M on the Conala and CodeAlpacaPy datasets (see Table~\ref{tab:rq2_table}), highlighting the clear advantages of PEFT in these scenarios.
However, due to resource constraints, we were unable to evaluate full fine-tuning for LLMs, leaving room for future studies to explore this further in the software engineering domain.

Additionally, our research uncovers new insights into the benefits of QLoRA and the comparative effectiveness of LoRA versus RAG for code generation tasks. 
First, in RQ3 and RQ4, we demonstrate that QLoRA offers comparable or even superior performance to LoRA while drastically cutting computational costs.
Second, we reveal limitations of ICL and RAG, showing that LLM effectiveness tends to plateau as more examples are retrieved.
In contrast, our study highlights the consistent advantages of PEFT techniques like LoRA and QLoRA in overcoming these limitations.

\subsubsection*{\textbf{SE tasks and multi-tasking}}
To ensure a focused study, we avoided adding extra tasks and datasets, preventing an excessively broad set of analyses. 
Exploring PEFT techniques for LLMs across varied tasks and datasets is a promising direction for future research. 
In particular, Lorahub~\cite{huang2023lorahub}, a recently introduced framework for multi-task learning, demonstrates that a composition of LoRA modules trained on different tasks can generalize to new, unseen tasks while offering a strong performance-efficiency trade-off. 
We believe applying similar approaches in AI for SE holds great potential, particularly as the research field aims at automating a broad range of code-related tasks.
\section{Threats to Validity}
\label{sec:threats}
\subsubsection*{\textbf{External validity}}
One main threat relates to the choice of our SLMs and LLMs. 
We mitigated this threat by carefully selecting a diverse set of models, as explained in Section~\ref{sec:baselines_llms}. 
These models encompass various families of LLMs, trained on distinct pre-training data and learning objectives, and varying in size.
Furthermore, we did not select larger model variants except when using QLoRA, as other PEFT techniques, ICL, and RAG limit the use of larger models within our resource constraints.

Another external threat to the validity is related to the quality and representativeness of the fine-tuning datasets.
To alleviate this concern, we chose the Conala dataset, which contains high-quality examples mined from StackOverflow posts. 
Additionally, this dataset has been representatively used by multiple prior studies~\cite{zhou2023docprompting,wang2022execution,norouzi2021code} on code generation tasks.
Furthermore, the authors enriched each natural language intent with hints, enhancing the alignment of input prompts with possible human intents.
To enrich our study, we included CodeAlpacaPy as a second dataset which encompasses lengthier examples, bringing another line of analysis. 
We did not include evaluation datasets such as HumanEval~\cite{chen2021evaluating} and MBPP~\cite{austin2021program}, as they do not include training examples.
However, to further expand our study, we explored the effectiveness of LoRA and QLoRA for execution-based code generation on the APPs dataset. 

Finally, the monolingual aspect of our datasets constitutes another threat to external validity.
We studied full fine-tuning, PEFT, ICL, and RAG for code generation of Python code snippets. 
However, we anticipate that PEFT is also applicable to other programming languages, considering the impressive generation capabilities of LLMs on a diverse range of programming languages~\cite{cassano2023multipl,athiwaratkun2022multi}.

\subsubsection*{\textbf{Internal validity}}
The hyperparameter choices for the PEFT methods constitute the main threat to internal validity. 
For each PEFT technique, we used hyperparameters values which have been used in previous work on PEFT for code models as well as in the seminal papers that contributed the PEFT techniques. 
Additionally, since LoRA with $r=16$ and $\alpha=32$ consistently outperforms all configurations of ICL and RAG across our top three models, conducting a detailed hyperparameter sensitivity analysis of LoRA could further solidify the advantage of PEFT over ICL and RAG. Future work could explore the sensitivity of key LoRA hyperparameters, such as rank $r$ and scaling factor $\alpha$, across a broader range of software engineering tasks.

\subsubsection*{\textbf{Construct validity}}
The choice of our evaluation metrics constitutes the main threat to construct validity.
To mitigate this threat, we selected evaluation metrics widely used in prior works~\cite{lu2021codexglue,wang2021codet5,yang2023exploitgen,li2023skcoder,apps2021,roziere2023code} on code generation. 
Furthermore, we evaluate each approach using EM@$k$ on Conala and CodeAlpacaPy, which enriched our analysis by computing the exact match over different ranges of code candidates.
Similarly, for APPs, we evaluate the base model and LoRA/QLoRA on Pass@$k$ with up to 5 candidates.
Finally, we did not use Pass@$k$ metrics as the CoNaLa and CodeAlpacaPy datasets do not include unit tests. 
Enriching the datasets with unit tests constitutes an interesting area of future work.
%yin2017syntactic,
\section{Related work}
\label{sec:related-work}

In this section, we overview existing work on LLMs for code generation and contrast previous contributions on efficient model adaptation of code for downstream tasks with our study.

\subsubsection*{Automated Code Generation}
A significant portion of code generation techniques~\cite{sun2020treegen,balog2016deepcoder,hayati2018retrieval,alon2020structural,wang2021codet5} relies on deep-learning-based approaches.
The latest trend in automated code generation revolves around leveraging LLMs like GPT models~\cite{openai2023gpt} due to their remarkable breakthroughs in this domain. 
One notable example is Codex, developed by Chen et al.~\cite{chen2021evaluating}, which is a fine-tuned version of GPT-3. 
Other noteworthy models following the success of Codex include CodeGen~\cite{nijkamp2023codegen}, CodeGen2~\cite{nijkamp2023codegen2} and CodeLlama~\cite{roziere2023code}. 
These LLMs effectively democratize the breakthrough performance achieved by Codex and bring it to a broader audience.
However, the high computational costs associated with full fine-tuning for LLMs to achieve optimal performance are impractical for most researchers and practitioners. 
We believe that our study can shed light on more efficient and cost-effective approaches to fine-tuning these LLMs, mitigating the computational burdens associated with their adoption.

\subsubsection*{Efficient Adaptation of Models of Code}
Efficient adaptation of models of code involves the utilization of techniques to efficiently adapt a model to a task-specific dataset (see Section~\ref{sec:background}). In this context, the term ``efficient'' refers to rendering the fine-tuning computation costs low, \textit{e.g}, using LoRA, or utilizing parameter-free techniques such as prompting and ICL.

Most prior research has concentrated on employing ICL and prompting to adapt models to diverse code-related tasks.
Gao et al.~\cite{gao2023constructing} showcased the advantages of ICL in tasks like bug fixing, code summarization, and program synthesis. 
They highlighted that the model's performance on downstream tasks is influenced by multiple factors, including the selection, quantity, and order of prompt examples. 
Other studies~\cite{prenner2022can, xia2023automated} also demonstrated that pre-trained language models and LLMs like Codex can effectively handle bug fixing and automated program repair using ICL.
Moreover, Geng et al.~\cite{geng2024large} demonstrated the capability of Codex to generate multi-intent comment generation to describe the functionality of a method or its implementation details, for instance.
The selection of relevant prompts for a task with ICL is crucial to ensure the good performance of an LLM. 
Prior works~\cite{nashid2023retrieval, zhou2023docprompting} designed selection techniques to retrieve highly relevant prompt examples tailored to downstream tasks, outperforming random selection methods.
Lastly, recent research~\cite{shrivastava2023repository} highlighted the advantages of retrieving prompt examples at the repository level, providing LLMs with valuable contextual information in the prompts.
In this study, we leveraged ICL without the intention of fully exploring its potential. 
Instead, we opted for a simple implementation of ICL by selecting random few-shot examples using different seeds.
Expanding this study to incorporate more ICL approaches would enhance the comparison with PEFT techniques for code.

Regarding PEFT techniques, prior research in code intelligence has focused on Prompt tuning~\cite{lester2021power}, Prefix-tuning~\cite{li2021prefix} and Adapters \cite{houlsby2019parameter,hu2023llm,goel2022cross,saberi2024utilization,saberi2023model}. 
Wang et al.~\cite{wang2022no} initiated the usage of Prompt tuning for code-related tasks and demonstrated its superiority over full fine-tuning of CodeT5 and CodeBERT in defect prediction, code summarization, and code translation. 
Goel et al.~\cite{goel2022cross} explored the use of programming-language-specific adapters for knowledge transfer in pre-trained language models, demonstrating that tuning BERT with these adapters surpass CodeBERT on cloze test and code clone detection.
Choi et al.~\cite{choi2023codeprompt} designed a code-specific Prefix tuning approach within a sequence-to-sequence architecture for generation tasks.
Our study differs from these three previous works as they focus on SLMs, whereas we propose the first comprehensive study of PEFT techniques with LLMs for code generation.
Moreover, our study includes LoRA, IA3, and QLoRA, which none of the previous work in code intelligence considered for efficiently tuning LLMs of code. 
Wang et al.~\cite{wang2023one} showcased the superiority of utilizing Adapters for fine-tuning pre-trained language models over full fine-tuning. 
Recent work have contributed empirical studies for various software engineering tasks, including code change~\cite{liu2024delving}, code summarization~\cite{saberi2024utilization, liu2023empirical}, defect prediction~\cite{liu2023empirical}, and code clone detection~\cite{saberi2024utilization}, using Adapter tuning and LoRA for SLMs.
Our research diverges from these prior work, as we concentrate on LLMs. 
Although we did not incorporate Adapters in our investigation, we believe that LoRA, IA3, Prompt tuning, Prefix tuning, and QLoRA provide a sufficiently thorough analysis of PEFT techniques.
We recognize the value of exploring additional PEFT techniques for various code intelligence tasks in the future.

\section{Conclusion and future work}
\label{sec:conclusion}

This study establishes the effectiveness of PEFT techniques in fine-tuning LLMs for code generation.
Our comparative analysis across various parameter-efficient techniques, including LoRA, IA3, Prompt tuning, Prefix tuning, and QLoRA, reveals the superiority of PEFT over full fine-tuning for SLMs and ICL and RAG for LLMs.
Furthermore, our study illustrates the practicality of PEFT under a limited resources scenario, effectively mitigating the reliance on large and expensive computational infrastructures.
To the best of our knowledge, this study is among the first comprehensive exploration of PEFT techniques for LLMs in software engineering, suggesting a promising avenue for future research.
We anticipate our findings will inspire further investigation into the application of PEFT techniques in software engineering, with potentially far-reaching impacts. 
Our future work will extend the study to alternative software engineering tasks such as automated code review and comment generation.
Finally, we aim to validate further relevance of PEFT techniques under multi-tasking and continual learning settings for automated software engineering.
% Finally, we aim at validating further reliance on PEFT techniques through continuously adapting LLMs to the larger scope.

% \begin{acks}
% \end{acks}

\bibliographystyle{ACM-Reference-Format}
\bibliography{references}


\begin{thebibliography}{92}

%%% the Web address.

\ifx \showCODEN    \undefined \def \showCODEN     #1{\unskip}     \fi
\ifx \showDOI      \undefined \def \showDOI       #1{#1}\fi
\ifx \showISBNx    \undefined \def \showISBNx     #1{\unskip}     \fi
\ifx \showISBNxiii \undefined \def \showISBNxiii  #1{\unskip}     \fi
\ifx \showISSN     \undefined \def \showISSN      #1{\unskip}     \fi
\ifx \showLCCN     \undefined \def \showLCCN      #1{\unskip}     \fi
\ifx \shownote     \undefined \def \shownote      #1{#1}          \fi
\ifx \showarticletitle \undefined \def \showarticletitle #1{#1}   \fi
\ifx \showURL      \undefined \def \showURL       {\relax}        \fi
% invisible to TeX
\providecommand\bibfield[2]{#2}
\providecommand\bibinfo[2]{#2}
\providecommand\natexlab[1]{#1}
\providecommand\showeprint[2][]{arXiv:#2}

\bibitem[Alon et~al\mbox{.}(2020)]%
        {alon2020structural}
\bibfield{author}{\bibinfo{person}{Uri Alon}, \bibinfo{person}{Roy Sadaka}, \bibinfo{person}{Omer Levy}, {and} \bibinfo{person}{Eran Yahav}.} \bibinfo{year}{2020}\natexlab{}.
\newblock \showarticletitle{Structural language models of code}. In \bibinfo{booktitle}{\emph{International conference on machine learning}}. PMLR, \bibinfo{pages}{245--256}.
\newblock


\bibitem[Athiwaratkun et~al\mbox{.}(2022)]%
        {athiwaratkun2022multi}
\bibfield{author}{\bibinfo{person}{Ben Athiwaratkun}, \bibinfo{person}{Sanjay~Krishna Gouda}, \bibinfo{person}{Zijian Wang}, \bibinfo{person}{Xiaopeng Li}, \bibinfo{person}{Yuchen Tian}, \bibinfo{person}{Ming Tan}, \bibinfo{person}{Wasi~Uddin Ahmad}, \bibinfo{person}{Shiqi Wang}, \bibinfo{person}{Qing Sun}, \bibinfo{person}{Mingyue Shang}, {et~al\mbox{.}}} \bibinfo{year}{2022}\natexlab{}.
\newblock \showarticletitle{Multi-lingual evaluation of code generation models}.
\newblock \bibinfo{journal}{\emph{arXiv preprint arXiv:2210.14868}} (\bibinfo{year}{2022}).
\newblock


\bibitem[Austin et~al\mbox{.}(2021)]%
        {austin2021program}
\bibfield{author}{\bibinfo{person}{Jacob Austin}, \bibinfo{person}{Augustus Odena}, \bibinfo{person}{Maxwell Nye}, \bibinfo{person}{Maarten Bosma}, \bibinfo{person}{Henryk Michalewski}, \bibinfo{person}{David Dohan}, \bibinfo{person}{Ellen Jiang}, \bibinfo{person}{Carrie Cai}, \bibinfo{person}{Michael Terry}, \bibinfo{person}{Quoc Le}, {et~al\mbox{.}}} \bibinfo{year}{2021}\natexlab{}.
\newblock \showarticletitle{Program synthesis with large language models}.
\newblock \bibinfo{journal}{\emph{arXiv preprint arXiv:2108.07732}} (\bibinfo{year}{2021}).
\newblock


\bibitem[Balog et~al\mbox{.}(2016)]%
        {balog2016deepcoder}
\bibfield{author}{\bibinfo{person}{Matej Balog}, \bibinfo{person}{Alexander~L Gaunt}, \bibinfo{person}{Marc Brockschmidt}, \bibinfo{person}{Sebastian Nowozin}, {and} \bibinfo{person}{Daniel Tarlow}.} \bibinfo{year}{2016}\natexlab{}.
\newblock \showarticletitle{Deepcoder: Learning to write programs}.
\newblock \bibinfo{journal}{\emph{arXiv preprint arXiv:1611.01989}} (\bibinfo{year}{2016}).
\newblock


\bibitem[Brown et~al\mbox{.}(2020)]%
        {brown2020language}
\bibfield{author}{\bibinfo{person}{Tom Brown}, \bibinfo{person}{Benjamin Mann}, \bibinfo{person}{Nick Ryder}, \bibinfo{person}{Melanie Subbiah}, \bibinfo{person}{Jared~D Kaplan}, \bibinfo{person}{Prafulla Dhariwal}, \bibinfo{person}{Arvind Neelakantan}, \bibinfo{person}{Pranav Shyam}, \bibinfo{person}{Girish Sastry}, \bibinfo{person}{Amanda Askell}, {et~al\mbox{.}}} \bibinfo{year}{2020}\natexlab{}.
\newblock \showarticletitle{Language models are few-shot learners}.
\newblock \bibinfo{journal}{\emph{Advances in neural information processing systems}}  \bibinfo{volume}{33} (\bibinfo{year}{2020}), \bibinfo{pages}{1877--1901}.
\newblock


\bibitem[Cassano et~al\mbox{.}(2023)]%
        {cassano2023multipl}
\bibfield{author}{\bibinfo{person}{Federico Cassano}, \bibinfo{person}{John Gouwar}, \bibinfo{person}{Daniel Nguyen}, \bibinfo{person}{Sydney Nguyen}, \bibinfo{person}{Luna Phipps-Costin}, \bibinfo{person}{Donald Pinckney}, \bibinfo{person}{Ming-Ho Yee}, \bibinfo{person}{Yangtian Zi}, \bibinfo{person}{Carolyn~Jane Anderson}, \bibinfo{person}{Molly~Q Feldman}, {et~al\mbox{.}}} \bibinfo{year}{2023}\natexlab{}.
\newblock \showarticletitle{MultiPL-E: a scalable and polyglot approach to benchmarking neural code generation}.
\newblock \bibinfo{journal}{\emph{IEEE Transactions on Software Engineering}} (\bibinfo{year}{2023}).
\newblock


\bibitem[Chappuis et~al\mbox{.}(2022)]%
        {chappuis2022prompt}
\bibfield{author}{\bibinfo{person}{Christel Chappuis}, \bibinfo{person}{Val{\'e}rie Zermatten}, \bibinfo{person}{Sylvain Lobry}, \bibinfo{person}{Bertrand Le~Saux}, {and} \bibinfo{person}{Devis Tuia}.} \bibinfo{year}{2022}\natexlab{}.
\newblock \showarticletitle{Prompt-RSVQA: Prompting visual context to a language model for remote sensing visual question answering}. In \bibinfo{booktitle}{\emph{Proceedings of the IEEE/CVF Conference on Computer Vision and Pattern Recognition}}. \bibinfo{pages}{1372--1381}.
\newblock


\bibitem[Chaudhary(2023)]%
        {CodeAlpacaDataset}
\bibfield{author}{\bibinfo{person}{Sahil Chaudhary}.} \bibinfo{year}{2023}\natexlab{}.
\newblock \bibinfo{title}{Code Alpaca: An Instruction-following LLaMA model for code generation}.
\newblock \bibinfo{howpublished}{\url{https://github.com/sahil280114/codealpaca}}.
\newblock


\bibitem[Chen et~al\mbox{.}(2021)]%
        {chen2021evaluating}
\bibfield{author}{\bibinfo{person}{Mark Chen}, \bibinfo{person}{Jerry Tworek}, \bibinfo{person}{Heewoo Jun}, \bibinfo{person}{Qiming Yuan}, \bibinfo{person}{Henrique Ponde de~Oliveira Pinto}, \bibinfo{person}{Jared Kaplan}, \bibinfo{person}{Harri Edwards}, \bibinfo{person}{Yuri Burda}, \bibinfo{person}{Nicholas Joseph}, \bibinfo{person}{Greg Brockman}, {et~al\mbox{.}}} \bibinfo{year}{2021}\natexlab{}.
\newblock \showarticletitle{Evaluating large language models trained on code}.
\newblock \bibinfo{journal}{\emph{arXiv preprint arXiv:2107.03374}} (\bibinfo{year}{2021}).
\newblock


\bibitem[Choi and Lee(2023)]%
        {choi2023codeprompt}
\bibfield{author}{\bibinfo{person}{YunSeok Choi} {and} \bibinfo{person}{Jee-Hyong Lee}.} \bibinfo{year}{2023}\natexlab{}.
\newblock \showarticletitle{CodePrompt: Task-Agnostic Prefix Tuning for Program and Language Generation}. In \bibinfo{booktitle}{\emph{Findings of the Association for Computational Linguistics: ACL 2023}}. \bibinfo{pages}{5282--5297}.
\newblock


\bibitem[Chowdhery et~al\mbox{.}(2022)]%
        {chowdhery2022palm}
\bibfield{author}{\bibinfo{person}{Aakanksha Chowdhery}, \bibinfo{person}{Sharan Narang}, \bibinfo{person}{Jacob Devlin}, \bibinfo{person}{Maarten Bosma}, \bibinfo{person}{Gaurav Mishra}, \bibinfo{person}{Adam Roberts}, \bibinfo{person}{Paul Barham}, \bibinfo{person}{Hyung~Won Chung}, \bibinfo{person}{Charles Sutton}, \bibinfo{person}{Sebastian Gehrmann}, {et~al\mbox{.}}} \bibinfo{year}{2022}\natexlab{}.
\newblock \showarticletitle{Palm: Scaling language modeling with pathways}.
\newblock \bibinfo{journal}{\emph{arXiv preprint arXiv:2204.02311}} (\bibinfo{year}{2022}).
\newblock


\bibitem[Dettmers et~al\mbox{.}(2022)]%
        {dettmers2022llmint8}
\bibfield{author}{\bibinfo{person}{Tim Dettmers}, \bibinfo{person}{Mike Lewis}, \bibinfo{person}{Younes Belkada}, {and} \bibinfo{person}{Luke Zettlemoyer}.} \bibinfo{year}{2022}\natexlab{}.
\newblock \showarticletitle{LLM.int8(): 8-bit Matrix Multiplication for Transformers at Scale}.
\newblock \bibinfo{journal}{\emph{arXiv preprint arXiv:2208.07339}} (\bibinfo{year}{2022}).
\newblock


\bibitem[Dettmers et~al\mbox{.}(2023)]%
        {dettmers2023qlora}
\bibfield{author}{\bibinfo{person}{Tim Dettmers}, \bibinfo{person}{Artidoro Pagnoni}, \bibinfo{person}{Ari Holtzman}, {and} \bibinfo{person}{Luke Zettlemoyer}.} \bibinfo{year}{2023}\natexlab{}.
\newblock \showarticletitle{Qlora: Efficient finetuning of quantized llms}.
\newblock \bibinfo{journal}{\emph{arXiv preprint arXiv:2305.14314}} (\bibinfo{year}{2023}).
\newblock


\bibitem[Ding et~al\mbox{.}(2022)]%
        {ding2022delta}
\bibfield{author}{\bibinfo{person}{Ning Ding}, \bibinfo{person}{Yujia Qin}, \bibinfo{person}{Guang Yang}, \bibinfo{person}{Fuchao Wei}, \bibinfo{person}{Zonghan Yang}, \bibinfo{person}{Yusheng Su}, \bibinfo{person}{Shengding Hu}, \bibinfo{person}{Yulin Chen}, \bibinfo{person}{Chi-Min Chan}, \bibinfo{person}{Weize Chen}, {et~al\mbox{.}}} \bibinfo{year}{2022}\natexlab{}.
\newblock \showarticletitle{Delta tuning: A comprehensive study of parameter efficient methods for pre-trained language models}.
\newblock \bibinfo{journal}{\emph{arXiv preprint arXiv:2203.06904}} (\bibinfo{year}{2022}).
\newblock


\bibitem[Ding et~al\mbox{.}(2023)]%
        {ding2023parameter}
\bibfield{author}{\bibinfo{person}{Ning Ding}, \bibinfo{person}{Yujia Qin}, \bibinfo{person}{Guang Yang}, \bibinfo{person}{Fuchao Wei}, \bibinfo{person}{Zonghan Yang}, \bibinfo{person}{Yusheng Su}, \bibinfo{person}{Shengding Hu}, \bibinfo{person}{Yulin Chen}, \bibinfo{person}{Chi-Min Chan}, \bibinfo{person}{Weize Chen}, {et~al\mbox{.}}} \bibinfo{year}{2023}\natexlab{}.
\newblock \showarticletitle{Parameter-efficient fine-tuning of large-scale pre-trained language models}.
\newblock \bibinfo{journal}{\emph{Nature Machine Intelligence}} \bibinfo{volume}{5}, \bibinfo{number}{3} (\bibinfo{year}{2023}), \bibinfo{pages}{220--235}.
\newblock


\bibitem[Feng et~al\mbox{.}(2020)]%
        {feng2020codebert}
\bibfield{author}{\bibinfo{person}{Zhangyin Feng}, \bibinfo{person}{Daya Guo}, \bibinfo{person}{Duyu Tang}, \bibinfo{person}{Nan Duan}, \bibinfo{person}{Xiaocheng Feng}, \bibinfo{person}{Ming Gong}, \bibinfo{person}{Linjun Shou}, \bibinfo{person}{Bing Qin}, \bibinfo{person}{Ting Liu}, \bibinfo{person}{Daxin Jiang}, {et~al\mbox{.}}} \bibinfo{year}{2020}\natexlab{}.
\newblock \showarticletitle{Codebert: A pre-trained model for programming and natural languages}.
\newblock \bibinfo{journal}{\emph{arXiv preprint arXiv:2002.08155}} (\bibinfo{year}{2020}).
\newblock


\bibitem[Gao et~al\mbox{.}(2023a)]%
        {gao2023constructing}
\bibfield{author}{\bibinfo{person}{Shuzheng Gao}, \bibinfo{person}{Xin-Cheng Wen}, \bibinfo{person}{Cuiyun Gao}, \bibinfo{person}{Wenxuan Wang}, {and} \bibinfo{person}{Michael~R Lyu}.} \bibinfo{year}{2023}\natexlab{a}.
\newblock \showarticletitle{Constructing Effective In-Context Demonstration for Code Intelligence Tasks: An Empirical Study}.
\newblock \bibinfo{journal}{\emph{arXiv preprint arXiv:2304.07575}} (\bibinfo{year}{2023}).
\newblock


\bibitem[Gao et~al\mbox{.}(2023b)]%
        {gao2023keeping}
\bibfield{author}{\bibinfo{person}{Shuzheng Gao}, \bibinfo{person}{Hongyu Zhang}, \bibinfo{person}{Cuiyun Gao}, {and} \bibinfo{person}{Chaozheng Wang}.} \bibinfo{year}{2023}\natexlab{b}.
\newblock \showarticletitle{Keeping Pace with Ever-Increasing Data: Towards Continual Learning of Code Intelligence Models}.
\newblock \bibinfo{journal}{\emph{arXiv preprint arXiv:2302.03482}} (\bibinfo{year}{2023}).
\newblock


\bibitem[Geng et~al\mbox{.}(2024)]%
        {geng2024large}
\bibfield{author}{\bibinfo{person}{Mingyang Geng}, \bibinfo{person}{Shangwen Wang}, \bibinfo{person}{Dezun Dong}, \bibinfo{person}{Haotian Wang}, \bibinfo{person}{Ge Li}, \bibinfo{person}{Zhi Jin}, \bibinfo{person}{Xiaoguang Mao}, {and} \bibinfo{person}{Xiangke Liao}.} \bibinfo{year}{2024}\natexlab{}.
\newblock \showarticletitle{Large Language Models are Few-Shot Summarizers: Multi-Intent Comment Generation via In-Context Learning}.
\newblock  (\bibinfo{year}{2024}).
\newblock


\bibitem[Goel et~al\mbox{.}(2022)]%
        {goel2022cross}
\bibfield{author}{\bibinfo{person}{Divyam Goel}, \bibinfo{person}{Ramansh Grover}, {and} \bibinfo{person}{Fatemeh~H Fard}.} \bibinfo{year}{2022}\natexlab{}.
\newblock \showarticletitle{On the cross-modal transfer from natural language to code through adapter modules}. In \bibinfo{booktitle}{\emph{Proceedings of the 30th IEEE/ACM International Conference on Program Comprehension}}. \bibinfo{pages}{71--81}.
\newblock


\bibitem[Hayati et~al\mbox{.}(2018)]%
        {hayati2018retrieval}
\bibfield{author}{\bibinfo{person}{Shirley~Anugrah Hayati}, \bibinfo{person}{Raphael Olivier}, \bibinfo{person}{Pravalika Avvaru}, \bibinfo{person}{Pengcheng Yin}, \bibinfo{person}{Anthony Tomasic}, {and} \bibinfo{person}{Graham Neubig}.} \bibinfo{year}{2018}\natexlab{}.
\newblock \showarticletitle{Retrieval-based neural code generation}.
\newblock \bibinfo{journal}{\emph{arXiv preprint arXiv:1808.10025}} (\bibinfo{year}{2018}).
\newblock


\bibitem[Hendrycks et~al\mbox{.}(2021)]%
        {apps2021}
\bibfield{author}{\bibinfo{person}{Dan Hendrycks}, \bibinfo{person}{Steven Basart}, \bibinfo{person}{Saurav Kadavath}, \bibinfo{person}{Mantas Mazeika}, \bibinfo{person}{Akul Arora}, \bibinfo{person}{Ethan Guo}, \bibinfo{person}{Collin Burns}, \bibinfo{person}{Samir Puranik}, \bibinfo{person}{Horace He}, \bibinfo{person}{Dawn Song}, {and} \bibinfo{person}{Jacob Steinhardt}.} \bibinfo{year}{2021}\natexlab{}.
\newblock \showarticletitle{Measuring Coding Challenge Competence With APPS}.
\newblock \bibinfo{journal}{\emph{NeurIPS}} (\bibinfo{year}{2021}).
\newblock


\bibitem[Houlsby et~al\mbox{.}(2019)]%
        {houlsby2019parameter}
\bibfield{author}{\bibinfo{person}{Neil Houlsby}, \bibinfo{person}{Andrei Giurgiu}, \bibinfo{person}{Stanislaw Jastrzebski}, \bibinfo{person}{Bruna Morrone}, \bibinfo{person}{Quentin De~Laroussilhe}, \bibinfo{person}{Andrea Gesmundo}, \bibinfo{person}{Mona Attariyan}, {and} \bibinfo{person}{Sylvain Gelly}.} \bibinfo{year}{2019}\natexlab{}.
\newblock \showarticletitle{Parameter-efficient transfer learning for NLP}. In \bibinfo{booktitle}{\emph{International Conference on Machine Learning}}. PMLR, \bibinfo{pages}{2790--2799}.
\newblock


\bibitem[Hu et~al\mbox{.}(2021)]%
        {hu2021lora}
\bibfield{author}{\bibinfo{person}{Edward~J Hu}, \bibinfo{person}{Yelong Shen}, \bibinfo{person}{Phillip Wallis}, \bibinfo{person}{Zeyuan Allen-Zhu}, \bibinfo{person}{Yuanzhi Li}, \bibinfo{person}{Shean Wang}, \bibinfo{person}{Lu Wang}, {and} \bibinfo{person}{Weizhu Chen}.} \bibinfo{year}{2021}\natexlab{}.
\newblock \showarticletitle{Lora: Low-rank adaptation of large language models}.
\newblock \bibinfo{journal}{\emph{arXiv preprint arXiv:2106.09685}} (\bibinfo{year}{2021}).
\newblock


\bibitem[Hu et~al\mbox{.}(2023)]%
        {hu2023llm}
\bibfield{author}{\bibinfo{person}{Zhiqiang Hu}, \bibinfo{person}{Yihuai Lan}, \bibinfo{person}{Lei Wang}, \bibinfo{person}{Wanyu Xu}, \bibinfo{person}{Ee-Peng Lim}, \bibinfo{person}{Roy Ka-Wei Lee}, \bibinfo{person}{Lidong Bing}, {and} \bibinfo{person}{Soujanya Poria}.} \bibinfo{year}{2023}\natexlab{}.
\newblock \showarticletitle{LLM-Adapters: An Adapter Family for Parameter-Efficient Fine-Tuning of Large Language Models}.
\newblock \bibinfo{journal}{\emph{arXiv preprint arXiv:2304.01933}} (\bibinfo{year}{2023}).
\newblock


\bibitem[Huang et~al\mbox{.}(2023)]%
        {huang2023lorahub}
\bibfield{author}{\bibinfo{person}{Chengsong Huang}, \bibinfo{person}{Qian Liu}, \bibinfo{person}{Bill~Yuchen Lin}, \bibinfo{person}{Tianyu Pang}, \bibinfo{person}{Chao Du}, {and} \bibinfo{person}{Min Lin}.} \bibinfo{year}{2023}\natexlab{}.
\newblock \showarticletitle{Lorahub: Efficient cross-task generalization via dynamic lora composition}.
\newblock \bibinfo{journal}{\emph{arXiv preprint arXiv:2307.13269}} (\bibinfo{year}{2023}).
\newblock


\bibitem[Joshi et~al\mbox{.}(2023)]%
        {joshi2023repair}
\bibfield{author}{\bibinfo{person}{Harshit Joshi}, \bibinfo{person}{Jos{\'e}~Cambronero Sanchez}, \bibinfo{person}{Sumit Gulwani}, \bibinfo{person}{Vu Le}, \bibinfo{person}{Gust Verbruggen}, {and} \bibinfo{person}{Ivan Radi{\v{c}}ek}.} \bibinfo{year}{2023}\natexlab{}.
\newblock \showarticletitle{Repair is nearly generation: Multilingual program repair with llms}. In \bibinfo{booktitle}{\emph{Proceedings of the AAAI Conference on Artificial Intelligence}}, Vol.~\bibinfo{volume}{37}. \bibinfo{pages}{5131--5140}.
\newblock


\bibitem[Kocetkov et~al\mbox{.}(2022)]%
        {Kocetkov2022TheStack}
\bibfield{author}{\bibinfo{person}{Denis Kocetkov}, \bibinfo{person}{Raymond Li}, \bibinfo{person}{Loubna Ben~Allal}, \bibinfo{person}{Jia Li}, \bibinfo{person}{Chenghao Mou}, \bibinfo{person}{Carlos Muñoz~Ferrandis}, \bibinfo{person}{Yacine Jernite}, \bibinfo{person}{Margaret Mitchell}, \bibinfo{person}{Sean Hughes}, \bibinfo{person}{Thomas Wolf}, \bibinfo{person}{Dzmitry Bahdanau}, \bibinfo{person}{Leandro von Werra}, {and} \bibinfo{person}{Harm de Vries}.} \bibinfo{year}{2022}\natexlab{}.
\newblock \showarticletitle{The Stack: 3 TB of permissively licensed source code}.
\newblock \bibinfo{journal}{\emph{Preprint}} (\bibinfo{year}{2022}).
\newblock


\bibitem[Kojima et~al\mbox{.}(2022)]%
        {kojima2022large}
\bibfield{author}{\bibinfo{person}{Takeshi Kojima}, \bibinfo{person}{Shixiang~Shane Gu}, \bibinfo{person}{Machel Reid}, \bibinfo{person}{Yutaka Matsuo}, {and} \bibinfo{person}{Yusuke Iwasawa}.} \bibinfo{year}{2022}\natexlab{}.
\newblock \showarticletitle{Large language models are zero-shot reasoners}.
\newblock \bibinfo{journal}{\emph{Advances in neural information processing systems}}  \bibinfo{volume}{35} (\bibinfo{year}{2022}), \bibinfo{pages}{22199--22213}.
\newblock


\bibitem[Lester et~al\mbox{.}(2021)]%
        {lester2021power}
\bibfield{author}{\bibinfo{person}{Brian Lester}, \bibinfo{person}{Rami Al-Rfou}, {and} \bibinfo{person}{Noah Constant}.} \bibinfo{year}{2021}\natexlab{}.
\newblock \showarticletitle{The Power of Scale for Parameter-Efficient Prompt Tuning}. In \bibinfo{booktitle}{\emph{Proceedings of the 2021 Conference on Empirical Methods in Natural Language Processing}}. \bibinfo{pages}{3045--3059}.
\newblock


\bibitem[Lewis et~al\mbox{.}(2020)]%
        {lewis2020retrieval}
\bibfield{author}{\bibinfo{person}{Patrick Lewis}, \bibinfo{person}{Ethan Perez}, \bibinfo{person}{Aleksandra Piktus}, \bibinfo{person}{Fabio Petroni}, \bibinfo{person}{Vladimir Karpukhin}, \bibinfo{person}{Naman Goyal}, \bibinfo{person}{Heinrich K{\"u}ttler}, \bibinfo{person}{Mike Lewis}, \bibinfo{person}{Wen-tau Yih}, \bibinfo{person}{Tim Rockt{\"a}schel}, {et~al\mbox{.}}} \bibinfo{year}{2020}\natexlab{}.
\newblock \showarticletitle{Retrieval-augmented generation for knowledge-intensive nlp tasks}.
\newblock \bibinfo{journal}{\emph{Advances in Neural Information Processing Systems}}  \bibinfo{volume}{33} (\bibinfo{year}{2020}), \bibinfo{pages}{9459--9474}.
\newblock


\bibitem[Li et~al\mbox{.}(2023a)]%
        {li2023skcoder}
\bibfield{author}{\bibinfo{person}{Jia Li}, \bibinfo{person}{Yongmin Li}, \bibinfo{person}{Ge Li}, \bibinfo{person}{Zhi Jin}, \bibinfo{person}{Yiyang Hao}, {and} \bibinfo{person}{Xing Hu}.} \bibinfo{year}{2023}\natexlab{a}.
\newblock \showarticletitle{Skcoder: A sketch-based approach for automatic code generation}.
\newblock \bibinfo{journal}{\emph{arXiv preprint arXiv:2302.06144}} (\bibinfo{year}{2023}).
\newblock


\bibitem[Li and Liang(2021)]%
        {li2021prefix}
\bibfield{author}{\bibinfo{person}{Xiang~Lisa Li} {and} \bibinfo{person}{Percy Liang}.} \bibinfo{year}{2021}\natexlab{}.
\newblock \showarticletitle{Prefix-tuning: Optimizing continuous prompts for generation}.
\newblock \bibinfo{journal}{\emph{arXiv preprint arXiv:2101.00190}} (\bibinfo{year}{2021}).
\newblock


\bibitem[Li et~al\mbox{.}(2023b)]%
        {li2023generaltextembeddingsmultistage}
\bibfield{author}{\bibinfo{person}{Zehan Li}, \bibinfo{person}{Xin Zhang}, \bibinfo{person}{Yanzhao Zhang}, \bibinfo{person}{Dingkun Long}, \bibinfo{person}{Pengjun Xie}, {and} \bibinfo{person}{Meishan Zhang}.} \bibinfo{year}{2023}\natexlab{b}.
\newblock \bibinfo{title}{Towards General Text Embeddings with Multi-stage Contrastive Learning}.
\newblock
\newblock
\showeprint[arxiv]{2308.03281}~[cs.CL]
\urldef\tempurl%
\url{https://arxiv.org/abs/2308.03281}
\showURL{%
\tempurl}


\bibitem[Liang et~al\mbox{.}(2022)]%
        {liang2022holistic}
\bibfield{author}{\bibinfo{person}{Percy Liang}, \bibinfo{person}{Rishi Bommasani}, \bibinfo{person}{Tony Lee}, \bibinfo{person}{Dimitris Tsipras}, \bibinfo{person}{Dilara Soylu}, \bibinfo{person}{Michihiro Yasunaga}, \bibinfo{person}{Yian Zhang}, \bibinfo{person}{Deepak Narayanan}, \bibinfo{person}{Yuhuai Wu}, \bibinfo{person}{Ananya Kumar}, {et~al\mbox{.}}} \bibinfo{year}{2022}\natexlab{}.
\newblock \showarticletitle{Holistic evaluation of language models}.
\newblock \bibinfo{journal}{\emph{arXiv preprint arXiv:2211.09110}} (\bibinfo{year}{2022}).
\newblock


\bibitem[Liang et~al\mbox{.}(2023)]%
        {liang2023gpt}
\bibfield{author}{\bibinfo{person}{W Liang}, \bibinfo{person}{M Yuksekgonul}, \bibinfo{person}{Y Mao}, \bibinfo{person}{E Wu}, {and} \bibinfo{person}{J Zou}.} \bibinfo{year}{2023}\natexlab{}.
\newblock \bibinfo{title}{GPT detectors are biased against non-native English writers (arXiv: 2304.02819). arXiv}.
\newblock
\newblock


\bibitem[Liu et~al\mbox{.}(2022)]%
        {liu2022few}
\bibfield{author}{\bibinfo{person}{Haokun Liu}, \bibinfo{person}{Derek Tam}, \bibinfo{person}{Mohammed Muqeeth}, \bibinfo{person}{Jay Mohta}, \bibinfo{person}{Tenghao Huang}, \bibinfo{person}{Mohit Bansal}, {and} \bibinfo{person}{Colin~A Raffel}.} \bibinfo{year}{2022}\natexlab{}.
\newblock \showarticletitle{Few-shot parameter-efficient fine-tuning is better and cheaper than in-context learning}.
\newblock \bibinfo{journal}{\emph{Advances in Neural Information Processing Systems}}  \bibinfo{volume}{35} (\bibinfo{year}{2022}), \bibinfo{pages}{1950--1965}.
\newblock


\bibitem[Liu et~al\mbox{.}(2023)]%
        {liu2023empirical}
\bibfield{author}{\bibinfo{person}{Jiaxing Liu}, \bibinfo{person}{Chaofeng Sha}, {and} \bibinfo{person}{Xin Peng}.} \bibinfo{year}{2023}\natexlab{}.
\newblock \showarticletitle{An Empirical Study of Parameter-Efficient Fine-Tuning Methods for Pre-Trained Code Models}. In \bibinfo{booktitle}{\emph{2023 38th IEEE/ACM International Conference on Automated Software Engineering (ASE)}}. IEEE, \bibinfo{pages}{397--408}.
\newblock


\bibitem[Liu et~al\mbox{.}(2020)]%
        {liu2020retrieval}
\bibfield{author}{\bibinfo{person}{Shangqing Liu}, \bibinfo{person}{Yu Chen}, \bibinfo{person}{Xiaofei Xie}, \bibinfo{person}{Jingkai Siow}, {and} \bibinfo{person}{Yang Liu}.} \bibinfo{year}{2020}\natexlab{}.
\newblock \showarticletitle{Retrieval-augmented generation for code summarization via hybrid gnn}.
\newblock \bibinfo{journal}{\emph{arXiv preprint arXiv:2006.05405}} (\bibinfo{year}{2020}).
\newblock


\bibitem[Liu et~al\mbox{.}(2024)]%
        {liu2024delving}
\bibfield{author}{\bibinfo{person}{Shuo Liu}, \bibinfo{person}{Jacky Keung}, \bibinfo{person}{Zhen Yang}, \bibinfo{person}{Fang Liu}, \bibinfo{person}{Qilin Zhou}, {and} \bibinfo{person}{Yihan Liao}.} \bibinfo{year}{2024}\natexlab{}.
\newblock \showarticletitle{Delving into Parameter-Efficient Fine-Tuning in Code Change Learning: An Empirical Study}.
\newblock \bibinfo{journal}{\emph{arXiv preprint arXiv:2402.06247}} (\bibinfo{year}{2024}).
\newblock


\bibitem[Lu et~al\mbox{.}(2022)]%
        {lu2022reacc}
\bibfield{author}{\bibinfo{person}{Shuai Lu}, \bibinfo{person}{Nan Duan}, \bibinfo{person}{Hojae Han}, \bibinfo{person}{Daya Guo}, \bibinfo{person}{Seung-won Hwang}, {and} \bibinfo{person}{Alexey Svyatkovskiy}.} \bibinfo{year}{2022}\natexlab{}.
\newblock \showarticletitle{Reacc: A retrieval-augmented code completion framework}.
\newblock \bibinfo{journal}{\emph{arXiv preprint arXiv:2203.07722}} (\bibinfo{year}{2022}).
\newblock


\bibitem[Lu et~al\mbox{.}(2021)]%
        {lu2021codexglue}
\bibfield{author}{\bibinfo{person}{Shuai Lu}, \bibinfo{person}{Daya Guo}, \bibinfo{person}{Shuo Ren}, \bibinfo{person}{Junjie Huang}, \bibinfo{person}{Alexey Svyatkovskiy}, \bibinfo{person}{Ambrosio Blanco}, \bibinfo{person}{Colin Clement}, \bibinfo{person}{Dawn Drain}, \bibinfo{person}{Daxin Jiang}, \bibinfo{person}{Duyu Tang}, {et~al\mbox{.}}} \bibinfo{year}{2021}\natexlab{}.
\newblock \showarticletitle{Codexglue: A machine learning benchmark dataset for code understanding and generation}.
\newblock \bibinfo{journal}{\emph{arXiv preprint arXiv:2102.04664}} (\bibinfo{year}{2021}).
\newblock


\bibitem[Mangrulkar et~al\mbox{.}(2022)]%
        {peft}
\bibfield{author}{\bibinfo{person}{Sourab Mangrulkar}, \bibinfo{person}{Sylvain Gugger}, \bibinfo{person}{Lysandre Debut}, \bibinfo{person}{Younes Belkada}, \bibinfo{person}{Sayak Paul}, {and} \bibinfo{person}{Benjamin Bossan}.} \bibinfo{year}{2022}\natexlab{}.
\newblock \bibinfo{title}{PEFT: State-of-the-art Parameter-Efficient Fine-Tuning methods}.
\newblock \bibinfo{howpublished}{\url{https://github.com/huggingface/peft}}.
\newblock


\bibitem[Min et~al\mbox{.}(2021b)]%
        {min2021recent}
\bibfield{author}{\bibinfo{person}{Bonan Min}, \bibinfo{person}{Hayley Ross}, \bibinfo{person}{Elior Sulem}, \bibinfo{person}{Amir Pouran~Ben Veyseh}, \bibinfo{person}{Thien~Huu Nguyen}, \bibinfo{person}{Oscar Sainz}, \bibinfo{person}{Eneko Agirre}, \bibinfo{person}{Ilana Heintz}, {and} \bibinfo{person}{Dan Roth}.} \bibinfo{year}{2021}\natexlab{b}.
\newblock \showarticletitle{Recent advances in natural language processing via large pre-trained language models: A survey}.
\newblock \bibinfo{journal}{\emph{Comput. Surveys}} (\bibinfo{year}{2021}).
\newblock


\bibitem[Min et~al\mbox{.}(2021a)]%
        {min2021metaicl}
\bibfield{author}{\bibinfo{person}{Sewon Min}, \bibinfo{person}{Mike Lewis}, \bibinfo{person}{Luke Zettlemoyer}, {and} \bibinfo{person}{Hannaneh Hajishirzi}.} \bibinfo{year}{2021}\natexlab{a}.
\newblock \showarticletitle{Metaicl: Learning to learn in context}.
\newblock \bibinfo{journal}{\emph{arXiv preprint arXiv:2110.15943}} (\bibinfo{year}{2021}).
\newblock


\bibitem[Nashid et~al\mbox{.}(2023)]%
        {nashid2023retrieval}
\bibfield{author}{\bibinfo{person}{Noor Nashid}, \bibinfo{person}{Mifta Sintaha}, {and} \bibinfo{person}{Ali Mesbah}.} \bibinfo{year}{2023}\natexlab{}.
\newblock \showarticletitle{Retrieval-based prompt selection for code-related few-shot learning}. In \bibinfo{booktitle}{\emph{Proceedings of the 45th International Conference on Software Engineering (ICSE’23)}}.
\newblock


\bibitem[Nijkamp et~al\mbox{.}(2023a)]%
        {nijkamp2023codegen2}
\bibfield{author}{\bibinfo{person}{Erik Nijkamp}, \bibinfo{person}{Hiroaki Hayashi}, \bibinfo{person}{Caiming Xiong}, \bibinfo{person}{Silvio Savarese}, {and} \bibinfo{person}{Yingbo Zhou}.} \bibinfo{year}{2023}\natexlab{a}.
\newblock \showarticletitle{Codegen2: Lessons for training llms on programming and natural languages}.
\newblock \bibinfo{journal}{\emph{arXiv preprint arXiv:2305.02309}} (\bibinfo{year}{2023}).
\newblock


\bibitem[Nijkamp et~al\mbox{.}(2023b)]%
        {nijkamp2023codegen}
\bibfield{author}{\bibinfo{person}{Erik Nijkamp}, \bibinfo{person}{Bo Pang}, \bibinfo{person}{Hiroaki Hayashi}, \bibinfo{person}{Lifu Tu}, \bibinfo{person}{Huan Wang}, \bibinfo{person}{Yingbo Zhou}, \bibinfo{person}{Silvio Savarese}, {and} \bibinfo{person}{Caiming Xiong}.} \bibinfo{year}{2023}\natexlab{b}.
\newblock \bibinfo{title}{CodeGen: An Open Large Language Model for Code with Multi-Turn Program Synthesis}.
\newblock
\newblock
\showeprint[arxiv]{2203.13474}~[cs.LG]


\bibitem[Norouzi et~al\mbox{.}(2021)]%
        {norouzi2021code}
\bibfield{author}{\bibinfo{person}{Sajad Norouzi}, \bibinfo{person}{Keyi Tang}, {and} \bibinfo{person}{Yanshuai Cao}.} \bibinfo{year}{2021}\natexlab{}.
\newblock \showarticletitle{Code generation from natural language with less prior knowledge and more monolingual data}. In \bibinfo{booktitle}{\emph{Proceedings of the 59th Annual Meeting of the Association for Computational Linguistics and the 11th International Joint Conference on Natural Language Processing (Volume 2: Short Papers)}}. \bibinfo{pages}{776--785}.
\newblock


\bibitem[OpenAI(2023)]%
        {openai2023gpt}
\bibfield{author}{\bibinfo{person}{R OpenAI}.} \bibinfo{year}{2023}\natexlab{}.
\newblock \showarticletitle{GPT-4 technical report}.
\newblock \bibinfo{journal}{\emph{arXiv}} (\bibinfo{year}{2023}), \bibinfo{pages}{2303--08774}.
\newblock


\bibitem[Ouyang et~al\mbox{.}(2022)]%
        {ouyang2022training}
\bibfield{author}{\bibinfo{person}{Long Ouyang}, \bibinfo{person}{Jeffrey Wu}, \bibinfo{person}{Xu Jiang}, \bibinfo{person}{Diogo Almeida}, \bibinfo{person}{Carroll Wainwright}, \bibinfo{person}{Pamela Mishkin}, \bibinfo{person}{Chong Zhang}, \bibinfo{person}{Sandhini Agarwal}, \bibinfo{person}{Katarina Slama}, \bibinfo{person}{Alex Ray}, {et~al\mbox{.}}} \bibinfo{year}{2022}\natexlab{}.
\newblock \showarticletitle{Training language models to follow instructions with human feedback}.
\newblock \bibinfo{journal}{\emph{Advances in Neural Information Processing Systems}}  \bibinfo{volume}{35} (\bibinfo{year}{2022}), \bibinfo{pages}{27730--27744}.
\newblock


\bibitem[Parvez et~al\mbox{.}(2021)]%
        {parvez2021retrieval}
\bibfield{author}{\bibinfo{person}{Md~Rizwan Parvez}, \bibinfo{person}{Wasi~Uddin Ahmad}, \bibinfo{person}{Saikat Chakraborty}, \bibinfo{person}{Baishakhi Ray}, {and} \bibinfo{person}{Kai-Wei Chang}.} \bibinfo{year}{2021}\natexlab{}.
\newblock \showarticletitle{Retrieval augmented code generation and summarization}.
\newblock \bibinfo{journal}{\emph{arXiv preprint arXiv:2108.11601}} (\bibinfo{year}{2021}).
\newblock


\bibitem[Prenner et~al\mbox{.}(2022)]%
        {prenner2022can}
\bibfield{author}{\bibinfo{person}{Julian~Aron Prenner}, \bibinfo{person}{Hlib Babii}, {and} \bibinfo{person}{Romain Robbes}.} \bibinfo{year}{2022}\natexlab{}.
\newblock \showarticletitle{Can OpenAI's codex fix bugs? an evaluation on QuixBugs}. In \bibinfo{booktitle}{\emph{Proceedings of the Third International Workshop on Automated Program Repair}}. \bibinfo{pages}{69--75}.
\newblock


\bibitem[Radford et~al\mbox{.}(2019)]%
        {radford2019language}
\bibfield{author}{\bibinfo{person}{Alec Radford}, \bibinfo{person}{Jeffrey Wu}, \bibinfo{person}{Rewon Child}, \bibinfo{person}{David Luan}, \bibinfo{person}{Dario Amodei}, \bibinfo{person}{Ilya Sutskever}, {et~al\mbox{.}}} \bibinfo{year}{2019}\natexlab{}.
\newblock \showarticletitle{Language models are unsupervised multitask learners}.
\newblock \bibinfo{journal}{\emph{OpenAI blog}} \bibinfo{volume}{1}, \bibinfo{number}{8} (\bibinfo{year}{2019}), \bibinfo{pages}{9}.
\newblock


\bibitem[Ren et~al\mbox{.}(2020)]%
        {ren2020codebleu}
\bibfield{author}{\bibinfo{person}{Shuo Ren}, \bibinfo{person}{Daya Guo}, \bibinfo{person}{Shuai Lu}, \bibinfo{person}{Long Zhou}, \bibinfo{person}{Shujie Liu}, \bibinfo{person}{Duyu Tang}, \bibinfo{person}{Neel Sundaresan}, \bibinfo{person}{Ming Zhou}, \bibinfo{person}{Ambrosio Blanco}, {and} \bibinfo{person}{Shuai Ma}.} \bibinfo{year}{2020}\natexlab{}.
\newblock \showarticletitle{Codebleu: a method for automatic evaluation of code synthesis}.
\newblock \bibinfo{journal}{\emph{arXiv preprint arXiv:2009.10297}} (\bibinfo{year}{2020}).
\newblock


\bibitem[Roziere et~al\mbox{.}(2023)]%
        {roziere2023code}
\bibfield{author}{\bibinfo{person}{Baptiste Roziere}, \bibinfo{person}{Jonas Gehring}, \bibinfo{person}{Fabian Gloeckle}, \bibinfo{person}{Sten Sootla}, \bibinfo{person}{Itai Gat}, \bibinfo{person}{Xiaoqing~Ellen Tan}, \bibinfo{person}{Yossi Adi}, \bibinfo{person}{Jingyu Liu}, \bibinfo{person}{Tal Remez}, \bibinfo{person}{J{\'e}r{\'e}my Rapin}, {et~al\mbox{.}}} \bibinfo{year}{2023}\natexlab{}.
\newblock \showarticletitle{Code llama: Open foundation models for code}.
\newblock \bibinfo{journal}{\emph{arXiv preprint arXiv:2308.12950}} (\bibinfo{year}{2023}).
\newblock


\bibitem[Saberi et~al\mbox{.}(2024)]%
        {saberi2024utilization}
\bibfield{author}{\bibinfo{person}{Iman Saberi}, \bibinfo{person}{Fatemeh Fard}, {and} \bibinfo{person}{Fuxiang Chen}.} \bibinfo{year}{2024}\natexlab{}.
\newblock \showarticletitle{Utilization of pre-trained language models for adapter-based knowledge transfer in software engineering}.
\newblock \bibinfo{journal}{\emph{Empirical Software Engineering}} \bibinfo{volume}{29}, \bibinfo{number}{4} (\bibinfo{year}{2024}), \bibinfo{pages}{94}.
\newblock


\bibitem[Saberi and Fard(2023)]%
        {saberi2023model}
\bibfield{author}{\bibinfo{person}{Iman Saberi} {and} \bibinfo{person}{Fatemeh~H Fard}.} \bibinfo{year}{2023}\natexlab{}.
\newblock \showarticletitle{Model-Agnostic Syntactical Information for Pre-Trained Programming Language Models}.
\newblock \bibinfo{journal}{\emph{arXiv preprint arXiv:2303.06233}} (\bibinfo{year}{2023}).
\newblock


\bibitem[Shao et~al\mbox{.}(2023)]%
        {shao2023prompting}
\bibfield{author}{\bibinfo{person}{Zhenwei Shao}, \bibinfo{person}{Zhou Yu}, \bibinfo{person}{Meng Wang}, {and} \bibinfo{person}{Jun Yu}.} \bibinfo{year}{2023}\natexlab{}.
\newblock \showarticletitle{Prompting large language models with answer heuristics for knowledge-based visual question answering}. In \bibinfo{booktitle}{\emph{Proceedings of the IEEE/CVF Conference on Computer Vision and Pattern Recognition}}. \bibinfo{pages}{14974--14983}.
\newblock


\bibitem[Shazeer and Stern(2018)]%
        {shazeer2018adafactor}
\bibfield{author}{\bibinfo{person}{Noam Shazeer} {and} \bibinfo{person}{Mitchell Stern}.} \bibinfo{year}{2018}\natexlab{}.
\newblock \showarticletitle{Adafactor: Adaptive learning rates with sublinear memory cost}. In \bibinfo{booktitle}{\emph{International Conference on Machine Learning}}. PMLR, \bibinfo{pages}{4596--4604}.
\newblock


\bibitem[Shrivastava et~al\mbox{.}(2023)]%
        {shrivastava2023repository}
\bibfield{author}{\bibinfo{person}{Disha Shrivastava}, \bibinfo{person}{Hugo Larochelle}, {and} \bibinfo{person}{Daniel Tarlow}.} \bibinfo{year}{2023}\natexlab{}.
\newblock \showarticletitle{Repository-level prompt generation for large language models of code}. In \bibinfo{booktitle}{\emph{International Conference on Machine Learning}}. PMLR, \bibinfo{pages}{31693--31715}.
\newblock


\bibitem[Strubell et~al\mbox{.}(2019)]%
        {strubell2019energy}
\bibfield{author}{\bibinfo{person}{Emma Strubell}, \bibinfo{person}{Ananya Ganesh}, {and} \bibinfo{person}{Andrew McCallum}.} \bibinfo{year}{2019}\natexlab{}.
\newblock \showarticletitle{Energy and policy considerations for deep learning in NLP}.
\newblock \bibinfo{journal}{\emph{arXiv preprint arXiv:1906.02243}} (\bibinfo{year}{2019}).
\newblock


\bibitem[Sun et~al\mbox{.}(2020)]%
        {sun2020treegen}
\bibfield{author}{\bibinfo{person}{Zeyu Sun}, \bibinfo{person}{Qihao Zhu}, \bibinfo{person}{Yingfei Xiong}, \bibinfo{person}{Yican Sun}, \bibinfo{person}{Lili Mou}, {and} \bibinfo{person}{Lu Zhang}.} \bibinfo{year}{2020}\natexlab{}.
\newblock \showarticletitle{Treegen: A tree-based transformer architecture for code generation}. In \bibinfo{booktitle}{\emph{Proceedings of the AAAI Conference on Artificial Intelligence}}, Vol.~\bibinfo{volume}{34}. \bibinfo{pages}{8984--8991}.
\newblock


\bibitem[Touvron et~al\mbox{.}(2023)]%
        {touvron2023llama}
\bibfield{author}{\bibinfo{person}{Hugo Touvron}, \bibinfo{person}{Thibaut Lavril}, \bibinfo{person}{Gautier Izacard}, \bibinfo{person}{Xavier Martinet}, \bibinfo{person}{Marie-Anne Lachaux}, \bibinfo{person}{Timoth{\'e}e Lacroix}, \bibinfo{person}{Baptiste Rozi{\`e}re}, \bibinfo{person}{Naman Goyal}, \bibinfo{person}{Eric Hambro}, \bibinfo{person}{Faisal Azhar}, {et~al\mbox{.}}} \bibinfo{year}{2023}\natexlab{}.
\newblock \showarticletitle{Llama: Open and efficient foundation language models}.
\newblock \bibinfo{journal}{\emph{arXiv preprint arXiv:2302.13971}} (\bibinfo{year}{2023}).
\newblock


\bibitem[Treviso et~al\mbox{.}(2023)]%
        {treviso2023efficient}
\bibfield{author}{\bibinfo{person}{Marcos Treviso}, \bibinfo{person}{Ji-Ung Lee}, \bibinfo{person}{Tianchu Ji}, \bibinfo{person}{Betty~van Aken}, \bibinfo{person}{Qingqing Cao}, \bibinfo{person}{Manuel~R Ciosici}, \bibinfo{person}{Michael Hassid}, \bibinfo{person}{Kenneth Heafield}, \bibinfo{person}{Sara Hooker}, \bibinfo{person}{Colin Raffel}, {et~al\mbox{.}}} \bibinfo{year}{2023}\natexlab{}.
\newblock \showarticletitle{Efficient methods for natural language processing: A survey}.
\newblock \bibinfo{journal}{\emph{Transactions of the Association for Computational Linguistics}}  \bibinfo{volume}{11} (\bibinfo{year}{2023}), \bibinfo{pages}{826--860}.
\newblock


\bibitem[Vaithilingam et~al\mbox{.}(2022)]%
        {vaithilingam2022expectation}
\bibfield{author}{\bibinfo{person}{Priyan Vaithilingam}, \bibinfo{person}{Tianyi Zhang}, {and} \bibinfo{person}{Elena~L Glassman}.} \bibinfo{year}{2022}\natexlab{}.
\newblock \showarticletitle{Expectation vs. experience: Evaluating the usability of code generation tools powered by large language models}. In \bibinfo{booktitle}{\emph{Chi conference on human factors in computing systems extended abstracts}}. \bibinfo{pages}{1--7}.
\newblock


\bibitem[Vaswani et~al\mbox{.}(2017)]%
        {vaswani2017attention}
\bibfield{author}{\bibinfo{person}{Ashish Vaswani}, \bibinfo{person}{Noam Shazeer}, \bibinfo{person}{Niki Parmar}, \bibinfo{person}{Jakob Uszkoreit}, \bibinfo{person}{Llion Jones}, \bibinfo{person}{Aidan~N Gomez}, \bibinfo{person}{{\L}ukasz Kaiser}, {and} \bibinfo{person}{Illia Polosukhin}.} \bibinfo{year}{2017}\natexlab{}.
\newblock \showarticletitle{Attention is all you need}.
\newblock \bibinfo{journal}{\emph{Advances in neural information processing systems}}  \bibinfo{volume}{30} (\bibinfo{year}{2017}).
\newblock


\bibitem[Wang et~al\mbox{.}(2022a)]%
        {wang2022no}
\bibfield{author}{\bibinfo{person}{Chaozheng Wang}, \bibinfo{person}{Yuanhang Yang}, \bibinfo{person}{Cuiyun Gao}, \bibinfo{person}{Yun Peng}, \bibinfo{person}{Hongyu Zhang}, {and} \bibinfo{person}{Michael~R Lyu}.} \bibinfo{year}{2022}\natexlab{a}.
\newblock \showarticletitle{No more fine-tuning? an experimental evaluation of prompt tuning in code intelligence}. In \bibinfo{booktitle}{\emph{Proceedings of the 30th ACM Joint European Software Engineering Conference and Symposium on the Foundations of Software Engineering}}. \bibinfo{pages}{382--394}.
\newblock


\bibitem[Wang et~al\mbox{.}(2023a)]%
        {wang2023one}
\bibfield{author}{\bibinfo{person}{Deze Wang}, \bibinfo{person}{Boxing Chen}, \bibinfo{person}{Shanshan Li}, \bibinfo{person}{Wei Luo}, \bibinfo{person}{Shaoliang Peng}, \bibinfo{person}{Wei Dong}, {and} \bibinfo{person}{Xiangke Liao}.} \bibinfo{year}{2023}\natexlab{a}.
\newblock \showarticletitle{One Adapter for All Programming Languages? Adapter Tuning for Code Search and Summarization}.
\newblock \bibinfo{journal}{\emph{arXiv preprint arXiv:2303.15822}} (\bibinfo{year}{2023}).
\newblock


\bibitem[Wang et~al\mbox{.}(2023c)]%
        {wang2023rap}
\bibfield{author}{\bibinfo{person}{Weishi Wang}, \bibinfo{person}{Yue Wang}, \bibinfo{person}{Shafiq Joty}, {and} \bibinfo{person}{Steven~CH Hoi}.} \bibinfo{year}{2023}\natexlab{c}.
\newblock \showarticletitle{Rap-gen: Retrieval-augmented patch generation with codet5 for automatic program repair}. In \bibinfo{booktitle}{\emph{Proceedings of the 31st ACM Joint European Software Engineering Conference and Symposium on the Foundations of Software Engineering}}. \bibinfo{pages}{146--158}.
\newblock


\bibitem[Wang et~al\mbox{.}(2023b)]%
        {wang2023codet5p}
\bibfield{author}{\bibinfo{person}{Yue Wang}, \bibinfo{person}{Hung Le}, \bibinfo{person}{Akhilesh~Deepak Gotmare}, \bibinfo{person}{Nghi~DQ Bui}, \bibinfo{person}{Junnan Li}, {and} \bibinfo{person}{Steven~CH Hoi}.} \bibinfo{year}{2023}\natexlab{b}.
\newblock \showarticletitle{Codet5+: Open code large language models for code understanding and generation}.
\newblock \bibinfo{journal}{\emph{arXiv preprint arXiv:2305.07922}} (\bibinfo{year}{2023}).
\newblock


\bibitem[Wang et~al\mbox{.}(2021)]%
        {wang2021codet5}
\bibfield{author}{\bibinfo{person}{Yue Wang}, \bibinfo{person}{Weishi Wang}, \bibinfo{person}{Shafiq Joty}, {and} \bibinfo{person}{Steven~CH Hoi}.} \bibinfo{year}{2021}\natexlab{}.
\newblock \showarticletitle{Codet5: Identifier-aware unified pre-trained encoder-decoder models for code understanding and generation}.
\newblock \bibinfo{journal}{\emph{arXiv preprint arXiv:2109.00859}} (\bibinfo{year}{2021}).
\newblock


\bibitem[Wang et~al\mbox{.}(2022b)]%
        {wang2022execution}
\bibfield{author}{\bibinfo{person}{Zhiruo Wang}, \bibinfo{person}{Shuyan Zhou}, \bibinfo{person}{Daniel Fried}, {and} \bibinfo{person}{Graham Neubig}.} \bibinfo{year}{2022}\natexlab{b}.
\newblock \showarticletitle{Execution-Based Evaluation for Open-Domain Code Generation}.
\newblock \bibinfo{journal}{\emph{arXiv preprint arXiv:2212.10481}} (\bibinfo{year}{2022}).
\newblock


\bibitem[Webson and Pavlick(2021)]%
        {webson2021prompt}
\bibfield{author}{\bibinfo{person}{Albert Webson} {and} \bibinfo{person}{Ellie Pavlick}.} \bibinfo{year}{2021}\natexlab{}.
\newblock \showarticletitle{Do prompt-based models really understand the meaning of their prompts?}
\newblock \bibinfo{journal}{\emph{arXiv preprint arXiv:2109.01247}} (\bibinfo{year}{2021}).
\newblock


\bibitem[Wei et~al\mbox{.}(2021)]%
        {wei2021finetuned}
\bibfield{author}{\bibinfo{person}{Jason Wei}, \bibinfo{person}{Maarten Bosma}, \bibinfo{person}{Vincent~Y Zhao}, \bibinfo{person}{Kelvin Guu}, \bibinfo{person}{Adams~Wei Yu}, \bibinfo{person}{Brian Lester}, \bibinfo{person}{Nan Du}, \bibinfo{person}{Andrew~M Dai}, {and} \bibinfo{person}{Quoc~V Le}.} \bibinfo{year}{2021}\natexlab{}.
\newblock \showarticletitle{Finetuned language models are zero-shot learners}.
\newblock \bibinfo{journal}{\emph{arXiv preprint arXiv:2109.01652}} (\bibinfo{year}{2021}).
\newblock


\bibitem[Wei et~al\mbox{.}(2022)]%
        {wei2022emergent}
\bibfield{author}{\bibinfo{person}{Jason Wei}, \bibinfo{person}{Yi Tay}, \bibinfo{person}{Rishi Bommasani}, \bibinfo{person}{Colin Raffel}, \bibinfo{person}{Barret Zoph}, \bibinfo{person}{Sebastian Borgeaud}, \bibinfo{person}{Dani Yogatama}, \bibinfo{person}{Maarten Bosma}, \bibinfo{person}{Denny Zhou}, \bibinfo{person}{Donald Metzler}, {et~al\mbox{.}}} \bibinfo{year}{2022}\natexlab{}.
\newblock \showarticletitle{Emergent abilities of large language models}.
\newblock \bibinfo{journal}{\emph{arXiv preprint arXiv:2206.07682}} (\bibinfo{year}{2022}).
\newblock


\bibitem[Weyssow et~al\mbox{.}(2022)]%
        {weyssow2022better}
\bibfield{author}{\bibinfo{person}{Martin Weyssow}, \bibinfo{person}{Houari Sahraoui}, {and} \bibinfo{person}{Bang Liu}.} \bibinfo{year}{2022}\natexlab{}.
\newblock \showarticletitle{Better modeling the programming world with code concept graphs-augmented multi-modal learning}. In \bibinfo{booktitle}{\emph{Proceedings of the ACM/IEEE 44th International Conference on Software Engineering: New Ideas and Emerging Results}}. \bibinfo{pages}{21--25}.
\newblock


\bibitem[Weyssow et~al\mbox{.}(2023)]%
        {weyssow2023usage}
\bibfield{author}{\bibinfo{person}{Martin Weyssow}, \bibinfo{person}{Xin Zhou}, \bibinfo{person}{Kisub Kim}, \bibinfo{person}{David Lo}, {and} \bibinfo{person}{Houari Sahraoui}.} \bibinfo{year}{2023}\natexlab{}.
\newblock \showarticletitle{On the Usage of Continual Learning for Out-of-Distribution Generalization in Pre-trained Language Models of Code}.
\newblock \bibinfo{journal}{\emph{arXiv preprint arXiv:2305.04106}} (\bibinfo{year}{2023}).
\newblock


\bibitem[Wolf et~al\mbox{.}(2019)]%
        {wolf2019huggingface}
\bibfield{author}{\bibinfo{person}{Thomas Wolf}, \bibinfo{person}{Lysandre Debut}, \bibinfo{person}{Victor Sanh}, \bibinfo{person}{Julien Chaumond}, \bibinfo{person}{Clement Delangue}, \bibinfo{person}{Anthony Moi}, \bibinfo{person}{Pierric Cistac}, \bibinfo{person}{Tim Rault}, \bibinfo{person}{R{\'e}mi Louf}, \bibinfo{person}{Morgan Funtowicz}, {et~al\mbox{.}}} \bibinfo{year}{2019}\natexlab{}.
\newblock \showarticletitle{Huggingface's transformers: State-of-the-art natural language processing}.
\newblock \bibinfo{journal}{\emph{arXiv preprint arXiv:1910.03771}} (\bibinfo{year}{2019}).
\newblock


\bibitem[Xia et~al\mbox{.}(2023)]%
        {xia2023automated}
\bibfield{author}{\bibinfo{person}{Chunqiu~Steven Xia}, \bibinfo{person}{Yuxiang Wei}, {and} \bibinfo{person}{Lingming Zhang}.} \bibinfo{year}{2023}\natexlab{}.
\newblock \showarticletitle{Automated program repair in the era of large pre-trained language models}. In \bibinfo{booktitle}{\emph{Proceedings of the 45th International Conference on Software Engineering (ICSE 2023). Association for Computing Machinery}}.
\newblock


\bibitem[Xia and Zhang(2022)]%
        {xia2022less}
\bibfield{author}{\bibinfo{person}{Chunqiu~Steven Xia} {and} \bibinfo{person}{Lingming Zhang}.} \bibinfo{year}{2022}\natexlab{}.
\newblock \showarticletitle{Less training, more repairing please: revisiting automated program repair via zero-shot learning}. In \bibinfo{booktitle}{\emph{Proceedings of the 30th ACM Joint European Software Engineering Conference and Symposium on the Foundations of Software Engineering}}. \bibinfo{pages}{959--971}.
\newblock


\bibitem[Xu et~al\mbox{.}(2022)]%
        {polycoder}
\bibfield{author}{\bibinfo{person}{Frank~F. Xu}, \bibinfo{person}{Uri Alon}, \bibinfo{person}{Graham Neubig}, {and} \bibinfo{person}{Vincent~Josua Hellendoorn}.} \bibinfo{year}{2022}\natexlab{}.
\newblock \showarticletitle{A Systematic Evaluation of Large Language Models of Code} \emph{(\bibinfo{series}{MAPS 2022})}. \bibinfo{publisher}{Association for Computing Machinery}, \bibinfo{address}{New York, NY, USA}, \bibinfo{pages}{1–10}.
\newblock
\showISBNx{9781450392730}
\urldef\tempurl%
\url{https://doi.org/10.1145/3520312.3534862}
\showDOI{\tempurl}


\bibitem[Yadav et~al\mbox{.}(2023)]%
        {yadav2023exploring}
\bibfield{author}{\bibinfo{person}{Prateek Yadav}, \bibinfo{person}{Qing Sun}, \bibinfo{person}{Hantian Ding}, \bibinfo{person}{Xiaopeng Li}, \bibinfo{person}{Dejiao Zhang}, \bibinfo{person}{Ming Tan}, \bibinfo{person}{Xiaofei Ma}, \bibinfo{person}{Parminder Bhatia}, \bibinfo{person}{Ramesh Nallapati}, \bibinfo{person}{Murali~Krishna Ramanathan}, {et~al\mbox{.}}} \bibinfo{year}{2023}\natexlab{}.
\newblock \showarticletitle{Exploring Continual Learning for Code Generation Models}.
\newblock \bibinfo{journal}{\emph{arXiv preprint arXiv:2307.02435}} (\bibinfo{year}{2023}).
\newblock


\bibitem[Yang et~al\mbox{.}(2023b)]%
        {yang2023exploitgen}
\bibfield{author}{\bibinfo{person}{Guang Yang}, \bibinfo{person}{Yu Zhou}, \bibinfo{person}{Xiang Chen}, \bibinfo{person}{Xiangyu Zhang}, \bibinfo{person}{Tingting Han}, {and} \bibinfo{person}{Taolue Chen}.} \bibinfo{year}{2023}\natexlab{b}.
\newblock \showarticletitle{ExploitGen: Template-augmented exploit code generation based on CodeBERT}.
\newblock \bibinfo{journal}{\emph{Journal of Systems and Software}}  \bibinfo{volume}{197} (\bibinfo{year}{2023}), \bibinfo{pages}{111577}.
\newblock


\bibitem[Yang et~al\mbox{.}(2023a)]%
        {yang2023language}
\bibfield{author}{\bibinfo{person}{Yue Yang}, \bibinfo{person}{Artemis Panagopoulou}, \bibinfo{person}{Shenghao Zhou}, \bibinfo{person}{Daniel Jin}, \bibinfo{person}{Chris Callison-Burch}, {and} \bibinfo{person}{Mark Yatskar}.} \bibinfo{year}{2023}\natexlab{a}.
\newblock \showarticletitle{Language in a bottle: Language model guided concept bottlenecks for interpretable image classification}. In \bibinfo{booktitle}{\emph{Proceedings of the IEEE/CVF Conference on Computer Vision and Pattern Recognition}}. \bibinfo{pages}{19187--19197}.
\newblock


\bibitem[Yin et~al\mbox{.}(2018a)]%
        {conala}
\bibfield{author}{\bibinfo{person}{Pengcheng Yin}, \bibinfo{person}{Bowen Deng}, \bibinfo{person}{Edgar Chen}, \bibinfo{person}{Bogdan Vasilescu}, {and} \bibinfo{person}{Graham Neubig}.} \bibinfo{year}{2018}\natexlab{a}.
\newblock \showarticletitle{Learning to Mine Aligned Code and Natural Language Pairs from Stack Overflow}. In \bibinfo{booktitle}{\emph{Proceedings of the 15th International Conference on Mining Software Repositories}} (Gothenburg, Sweden) \emph{(\bibinfo{series}{MSR '18})}. \bibinfo{publisher}{Association for Computing Machinery}, \bibinfo{address}{New York, NY, USA}, \bibinfo{pages}{476–486}.
\newblock
\showISBNx{9781450357166}
\urldef\tempurl%
\url{https://doi.org/10.1145/3196398.3196408}
\showDOI{\tempurl}


\bibitem[Yin et~al\mbox{.}(2018b)]%
        {yin2018learning}
\bibfield{author}{\bibinfo{person}{Pengcheng Yin}, \bibinfo{person}{Bowen Deng}, \bibinfo{person}{Edgar Chen}, \bibinfo{person}{Bogdan Vasilescu}, {and} \bibinfo{person}{Graham Neubig}.} \bibinfo{year}{2018}\natexlab{b}.
\newblock \showarticletitle{Learning to mine aligned code and natural language pairs from stack overflow}. In \bibinfo{booktitle}{\emph{2018 IEEE/ACM 15th international conference on mining software repositories (MSR)}}. IEEE, \bibinfo{pages}{476--486}.
\newblock


\bibitem[Yuan et~al\mbox{.}(2023)]%
        {yuan2023evaluating}
\bibfield{author}{\bibinfo{person}{Zhiqiang Yuan}, \bibinfo{person}{Junwei Liu}, \bibinfo{person}{Qiancheng Zi}, \bibinfo{person}{Mingwei Liu}, \bibinfo{person}{Xin Peng}, {and} \bibinfo{person}{Yiling Lou}.} \bibinfo{year}{2023}\natexlab{}.
\newblock \showarticletitle{Evaluating instruction-tuned large language models on code comprehension and generation}.
\newblock \bibinfo{journal}{\emph{arXiv preprint arXiv:2308.01240}} (\bibinfo{year}{2023}).
\newblock


\bibitem[Zhang et~al\mbox{.}(2023)]%
        {zhang2023investigating}
\bibfield{author}{\bibinfo{person}{Bowen Zhang}, \bibinfo{person}{Xianghua Fu}, \bibinfo{person}{Daijun Ding}, \bibinfo{person}{Hu Huang}, \bibinfo{person}{Yangyang Li}, {and} \bibinfo{person}{Liwen Jing}.} \bibinfo{year}{2023}\natexlab{}.
\newblock \showarticletitle{Investigating Chain-of-thought with ChatGPT for Stance Detection on Social Media}.
\newblock \bibinfo{journal}{\emph{arXiv preprint arXiv:2304.03087}} (\bibinfo{year}{2023}).
\newblock


\bibitem[Zhao et~al\mbox{.}(2021)]%
        {zhao2021calibrate}
\bibfield{author}{\bibinfo{person}{Zihao Zhao}, \bibinfo{person}{Eric Wallace}, \bibinfo{person}{Shi Feng}, \bibinfo{person}{Dan Klein}, {and} \bibinfo{person}{Sameer Singh}.} \bibinfo{year}{2021}\natexlab{}.
\newblock \showarticletitle{Calibrate before use: Improving few-shot performance of language models}. In \bibinfo{booktitle}{\emph{International Conference on Machine Learning}}. PMLR, \bibinfo{pages}{12697--12706}.
\newblock


\bibitem[Zhou et~al\mbox{.}(2023)]%
        {zhou2023docprompting}
\bibfield{author}{\bibinfo{person}{Shuyan Zhou}, \bibinfo{person}{Uri Alon}, \bibinfo{person}{Frank~F Xu}, \bibinfo{person}{Zhengbao Jiang}, {and} \bibinfo{person}{Graham Neubig}.} \bibinfo{year}{2023}\natexlab{}.
\newblock \showarticletitle{Docprompting: Generating code by retrieving the docs}. In \bibinfo{booktitle}{\emph{The Eleventh International Conference on Learning Representations}}.
\newblock


\bibitem[Zhou et~al\mbox{.}(2021)]%
        {zhou2021assessing}
\bibfield{author}{\bibinfo{person}{Xin Zhou}, \bibinfo{person}{DongGyun Han}, {and} \bibinfo{person}{David Lo}.} \bibinfo{year}{2021}\natexlab{}.
\newblock \showarticletitle{Assessing generalizability of codebert}. In \bibinfo{booktitle}{\emph{2021 IEEE International Conference on Software Maintenance and Evolution (ICSME)}}. IEEE, \bibinfo{pages}{425--436}.
\newblock


\end{thebibliography}

% \appendix

\end{document}